\documentclass[twocolumn]{aastex61}

\usepackage{epstopdf}
\usepackage{verbatim}
\usepackage{natbib}
\usepackage{apjfonts}
\usepackage{hyperref}
\usepackage{float}

\def\lae{\mathrel{<\kern-1.0em\lower0.9ex\hbox{$\sim$}}}
\def\gae{\mathrel{>\kern-1.0em\lower0.9ex\hbox{$\sim$}}}

\begin{document}

\shortauthors{Breiding et al.}{}
\shorttitle{Fermi non-detections challenge the IC/CMB Model}

\title{Fermi non-detections of four X-ray Jet Sources and Implications for the IC/CMB Mechanism}

\author{Peter Breiding}   
\affil{Department of Physics, University of Maryland Baltimore County, Baltimore, MD 21250, USA}
\author{Eileen T. Meyer}
\affil{Department of Physics, University of Maryland Baltimore County, Baltimore, MD 21250, USA}
\correspondingauthor{Eileen T. Meyer}
\email{meyer@umbc.edu}
\author{Markos Georganopoulos}
\affil{Department of Physics, University of Maryland Baltimore County, Baltimore, MD 21250, USA}
\affil{NASA Goddard Space Flight Center, Code 663, Greenbelt, MD 20771, USA}
\author{M. E. Keenan}
\affil{Department of Physics, University of Maryland Baltimore County, Baltimore, MD 21250, USA}
\author{N. S. DeNigris}
\affil{Department of Physics, University of Maryland Baltimore County, Baltimore, MD 21250, USA}
\author{Jennifer Hewitt}
\affil{Department of Physics, University of Maryland Baltimore County, Baltimore, MD 21250, USA}
\begin{abstract}
Since its launch in 1999, the \textit{Chandra} X-ray observatory has
discovered several dozen X-ray jets
associated with powerful quasars.  In many cases the X-ray spectrum is hard and appears to come from a second spectral component.  The most popular explanation for
the kpc-scale X-ray emission in these cases has been inverse-Compton
(IC) scattering of Cosmic Microwave Background (CMB) photons by
relativistic electrons in the jet (the IC/CMB model). Requiring the
IC/CMB emission to reproduce the observed X-ray flux density inevitably
predicts a high level of gamma-ray emission which should be detectable
with the \textit{Fermi} Large Area Telescope (LAT).  In previous work,
we found that gamma-ray upper limits from the large scale
jets of 3C~273 and PKS~0637$-$752 violate the predictions of the
IC/CMB model.  Here we present \textit{Fermi}/LAT flux density upper limits
for the X-ray jets of four additional sources: PKS~1136-135, PKS~1229-021, PKS~1354+195, and PKS~2209+080, and show that these
limits violate the IC/CMB predictions at a very high significance
level.  We also present new Hubble Space Telescope (HST) observations
of the quasar PKS~2209+080 showing a newly detected optical jet, and
Atacama Large Millimeter/submillimeter Array (ALMA) band 3 and 6
observations of all four sources, which provide key constraints on the
spectral shape that enable us to rule out the IC/CMB model.
\end{abstract}



\section{Introduction}

The first astrophysical target observed by the \emph{Chandra} X-ray Observatory was
PKS~0637-752, a relatively distant (z = 0.651) radio-loud active
galactic nucleus (AGN).  Unexpectedly, \textit{Chandra} discovered high
levels of X-ray emission associated with the previously known
kpc-scale radio jet \citep{schwartz2000,chartas2000}.  The
\textit{Chandra} observations revealed bright X-ray knots roughly
co-spatial with the radio knots, but with an X-ray flux density far too
high and an X-ray spectrum far too hard to be consistent with a single radio-optical-X-ray
synchrotron spectrum.  In the years that followed, this phenomenon of
bright X-ray jets in which the X-ray emission could not be explained
by an extrapolation of the radio-optical synchrotron spectrum
was found to operate in dozens of other sources, nearly always in more
powerful jets with a Fanaroff \& Riley (FR) type II
morphology \citep[see][ for a review]{harris2006}\footnote{Here we use the term FR to refer to the
  Fanaroff-Riley classification of radio galaxies, where FR~I-type
  sources are of lower radio luminosity and higher surface
  brightnesses towards the core than the FR~II-type sources which
  typically terminate in bright "hot spots"
  \citep{fanaroffriley}.}.
In this paper, we refer to any jet with evidence of multiple spectra as a Multiple Spectral Component or "MSC" jet.  Typically the observed X-ray emission in less powerful FR~I jets is more consistent
with a single radio to X-ray synchrotron spectrum
\citep[e.g.][]{wilson2002}

In the original papers announcing the detection of the MSC jet associated with PKS~0637-752 by
\textit{Chandra}, \cite{chartas2000} and \cite{schwartz2000} ruled out
thermal bremsstrahlung, synchrotron self-Compton, and inverse Compton
off the Cosmic Microwave Background (IC/CMB) from a mildly
relativistic kpc-scale jet as the possible X-ray emission
mechanisms.  However, proper motion
studies of PKS~0637$-$752 with Very Long Baseline Interferometry
(VLBI) had shown highly superluminal apparent velocities of $11.4\pm
0.6c$ on parsec scales, suggesting a highly relativistic flow
($\Gamma>11.4$) and a jet angle to the line-of-sight less than
$8.9^{\circ}$ \citep{lovell2000}. Independently, both
\cite{tavecchio2000} and \cite{celotti2001} suggested that the IC/CMB
model could produce the required X-ray flux density in PKS~0637$-$752 if the
jet remained highly relativistic on kpc scales ($\Gamma\sim10$ rather
than $\Gamma\sim1-2$) and was pointed at a small angle with respect to
the line-of-sight.  This model has been widely adopted by the
community as the probable X-ray emission mechanism for MSC
jets \citep[e.g.][]{sambruna2004, kharb2012,jorstad2006,
  marshall2011, stanley15, miller06,tavecchio2007, perlman11,
  sambruna02}.
While many AGN jets have been
shown to be highly relativistic on the sub-pc scale via proper motion studies with VLBI \citep[e.g.,][]{jorstad2005,lister2009},
this is no guarantee they remain so on the kpc scale, with
population-based evidence suggesting they are at most mildly
relativistic (\citealt{mullin2009} found $\Gamma\approx1.18-1.49$ for FR~II sources; also see \citealt{arshakian2004}).
    
\subsection{Problems with the IC/CMB model}
    
  
 One problematic aspect of the IC/CMB model is that it is energetically
 costly, with kinetic jet powers at or sometimes exceeding the
 Eddington luminosity for these sources \citep{dermer2004}.  This is
 primarily because it requires extending the electron energy
 distribution to very low energies to provide electrons of the right energy to
 produce the observed X-ray emission from upscattered infrared CMB
 photons.  This low-energy extension (i.e., a low
 $\gamma_\mathrm{min}$ value) of the electron energy distribution
 greatly increases the total energy content of the jet by requiring a
 much larger total number of electrons.  While such an extension is
 possible, there is no direct evidence for these low energy electrons
 since their synchrotron radiation would be at frequencies much less
 than our GHz-frequency radio observations.

 The radiative losses of the electrons responsible for the X-ray
 emission are very weak in the IC/CMB model, with radiative lifetimes
 in excess of $10^{6}$ years \citep{harris2006}.  In the simplest
 case, we would thus expect continuous X-ray emission along the
 entire jet length rather than the observed X-ray knots as noted by \cite{atoyan2004} and \cite{stawarz2004}.  This problem can be avoided
 if the X-ray emitting plasma is confined to moving blobs rather than
 a continuous jet flow \citep{tavecchio03}.  This solution would
 require these X-ray emitting blobs to be propagating with a high
 $\Gamma$ which may be observable with proper motion studies for
 nearby sources.  However, in the case of 3C 273, \cite{meyer16_hst}
 found that the knots are stationary with a limit of apparent speed
 $\beta_{app}\ <\ 1c$, corresponding to a limit of $\Gamma\ <\ 2.9$,
 which is incompatible with the IC/CMB interpretation for the X-rays.

 A feature of some of these MSC jets is offsets in peak
 brightness between radio, optical, and X-ray wavelengths when
 observed with similar angular resolutions (see the case of 3C 111 in
 \citealt{clautice2016} for a clear example of this).  There is no
 reasonable explanation for these offsets in the IC/CMB model since
 they are inconsistent with the idea that the X-ray, radio, and
 optical emission is produced by the same population of relativistic
 electrons, where we would expect co-spatial emission regions.

In addition to the above difficulties, there have been several
observations of MSC jets which directly challenge the
IC/CMB model.  \textit{HST} polarization measurements of the rising
UV-component in the jet of PKS~1136-135, clearly part of the second
spectral component, show fractional polarization measures in excess of
30\% for several knots \citep{cara2013}.  Since IC/CMB emission is
expected to have very low polarization (see \citealt{McNamara2009},
\citealt{uchiyama2008}), it is extremely unlikely that the second
component is produced by the IC/CMB mechanism in this source.
Another recent study by \cite{Hardcastle2016} of the nearby FR~II
radio galaxy Pictor A shows IC/CMB to be incompatible with the
observed jet to counter-jet flux ratio.  They found a ratio orders of
magnitude less than what would be expected if IC/CMB were the source
of the X-rays (in which case $\Gamma\geq5$ and a jet angle less than a
few degrees is required).  While the conclusions drawn from these observations of PKS~1136-135,
3C~273, and Pictor A do not necessarily apply to all MSC
jets, they already necessitate a second emission mechanism which is
not IC/CMB to explain the X-ray emission in these cases.

There are several alternative models which might explain the MSC jets.  In
particular, a second population of synchrotron-emitting electrons
(\citealt{atoyan2004}; \citealt{harris2004}; \citealt{kataoka2005};
\citealt{hardcastle2006}; \citealt{jester2006};
\citealt{uchiyama2006}) and various hadronic emission models
(\citealt{aharonian2002}; \citealt{petropoulou2017};
\citealt{kusunose17}).  Until recently it has been difficult to reject
the IC/CMB model in favor of any alternative since in most cases any model can be tuned to reproduce the somewhat sparsely sampled spectral energy
distribution (SED).  However, the question of which emission process
is at work in these X-ray jets is important since these models imply
vastly different physical properties.  

In the case of the second synchrotron model, it will be important to understand why
there are multiple electron energy distributions and how this is
connected to the particle acceleration taking place in these jets.  An
immediate consequence of a higher-energy electron distribution is
subsequent TeV gamma-rays produced when these electrons
inverse-Compton scatter the CMB (\citealt{georganopoulos2006};
\citealt{meyer2015}). These TeV gamma-rays may be detectable for low
redshift sources with the upcoming Cherenkov Telescope Array
(\textit{CTA}).  Alternatively, if the source of the X-rays is due to
hadronic emission processes then this has important implications for
the particle make-up and energy content of these jets.


\subsection{Using the Fermi/LAT to test the IC/CMB model}

	 As noted by \cite{georganopoulos2006}, the IC/CMB model
         predicts a high level of IC/CMB gamma-ray emission from the
         jet which should be detectable with the \textit{Fermi}/LAT.
         In previous work, we developed a technique to search for the
         steady gamma-ray IC/CMB emission component at times
         when the variable core was in a low state, necessary since
         the resolution of \textit{Fermi} is insufficient to
         separately resolve the core and jet.  This technique was
         applied in 3C 273 \citep{meyer2014} and PKS 0637-752
         (\citealt{meyer2015}; \citealt{meyer17}), where in both cases the upper limits to the gamma-ray flux during times when the core was quiescent were well
         below the required level from IC/CMB.  This paper follows
         these works for four more MSC jets where the
         X-ray emission has been previously modeled as due to IC/CMB.
         However, in this paper we have targeted sources which are not detected in the \textit{Fermi}/LAT 3FGL 4-year point
         source catalog, which strongly suggests that the gamma-ray
         emission expected under the IC/CMB model may be absent,
         without any complication of gamma-ray emission from the core.  Additional MSC jets are currently being analyzed in order to establish which sources may be consistent with the IC/CMB model based on gamma-ray observations, but the presence of newly detected point sources in the region of interest makes the analysis more complicated and we leave the discussion of these sources to a future paper.

         In Section~\ref{sec_data} we present the new and archival
         radio and optical data analyzed for our four targets, as well
         as the method of deriving \emph{Fermi} upper limits. In
         Section~\ref{sec_results} we present the resulting limits
         which clearly violate the expectations of the IC/CMB model.  In Section~\ref{sec_disc} we discuss the properties of the
         small but growing sample of MSC jets in which IC/CMB
         has been ruled out, including several lines of evidence that
         suggest that they are not extremely well-aligned as required
         under the IC/CMB model.  Finally, in Section~\ref{sec_summ}
         we summarize our findings.

\section{Data Analysis}
\label{sec_data}
\subsection{Sample Properties}

\begin{figure*}[ht]
	\vspace{20pt}
	\begin{center}
		\includegraphics[width=6in]{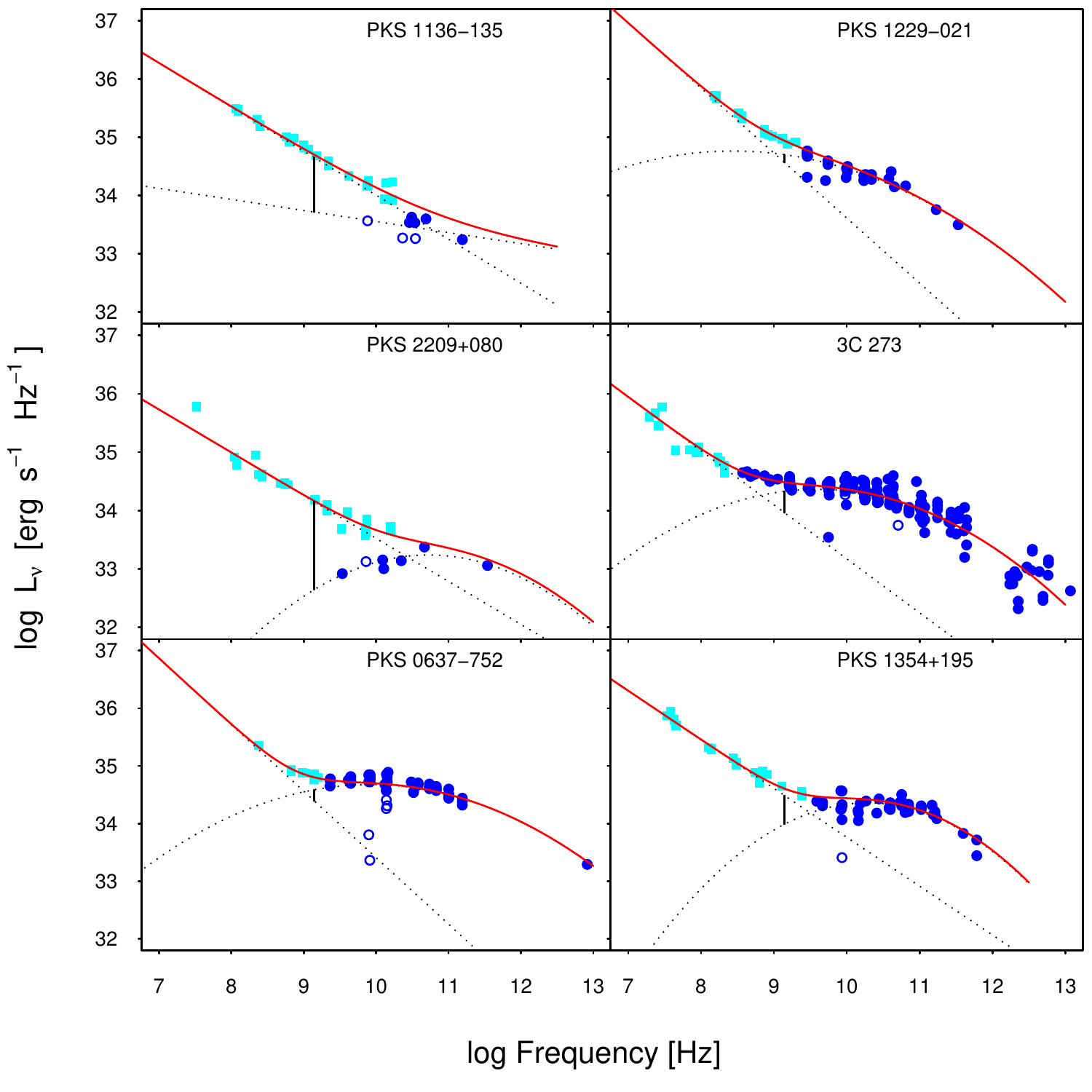}
	\end{center}
	\caption{\label{fig:radio seds}Low frequency radio SEDs for the six sources where we have now ruled out IC/CMB as the X-ray emission mechanism.  Data was obtained through the NED database, cyan squares represent extended lobe emission, blue circles represent core emission, and open circles represent data filtered out from the fitting.  A power-law model was used to fit the lobe emission and a log parabola was used to fit the core emission, dashed black lines showing each of the individual fits.  The solid red line shows the total model fit adding both of these components for the radio spectrum.  The solid black line connects the core and extended emission at 1.4~GHz, the ratio of which was used to determine the radio core dominance.}
\end{figure*}

\begin{figure*}[ht]
	\vspace{20pt}
	\begin{center}
		\includegraphics[width=6in]{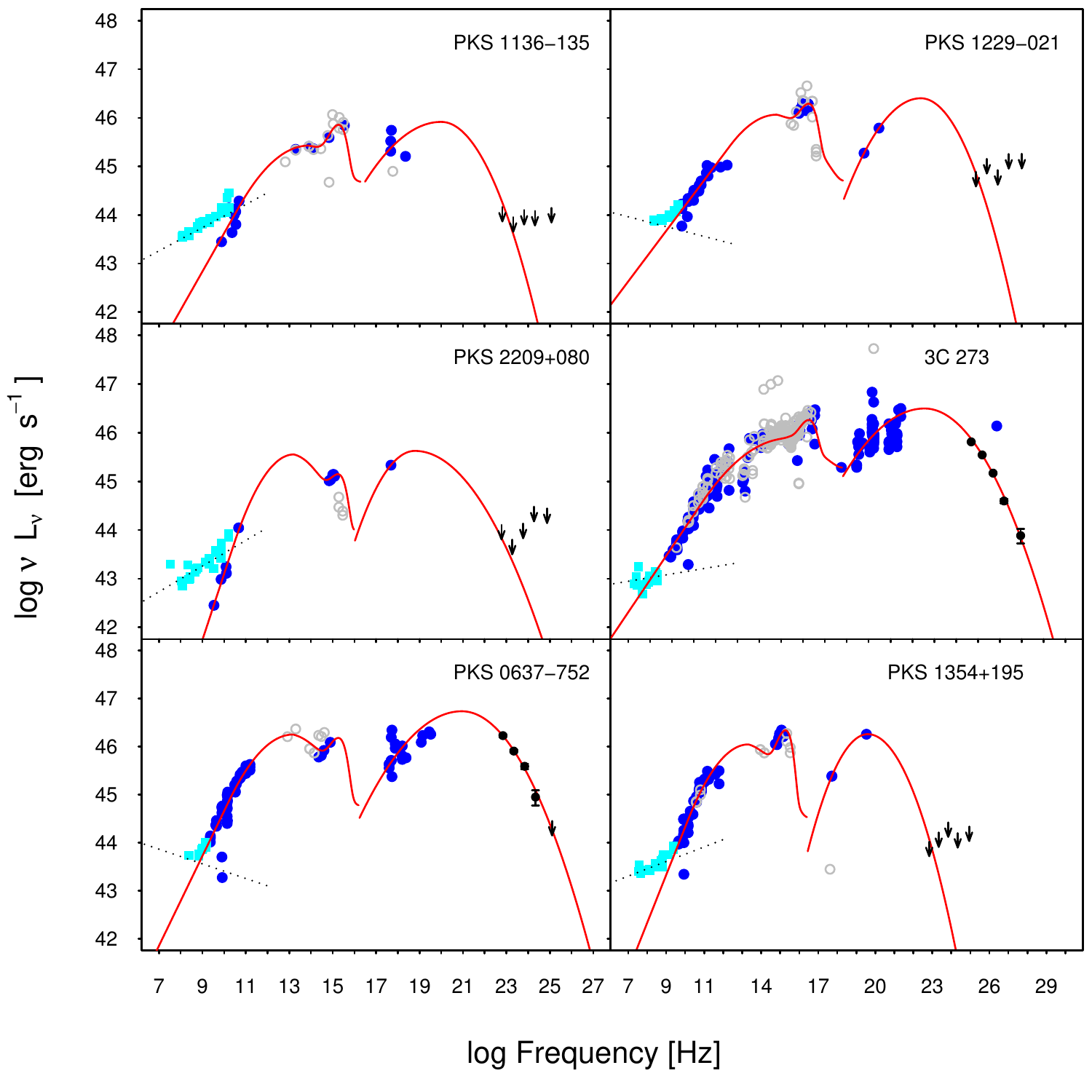}
	\end{center}
	\caption{\label{fig:source seds} The SED for each source was fit with the parametric model from \cite{meyer2011} shown in red to estimate parameters for the synchrotron peak frequencies and luminosities. All data was obtained through NED, with the cyan squares representing the lobe spectrum with the line of best fit as the dotted black line.  Blue circles represent the rest of data and gray open circles represent the data that was filtered out and not used in the fitting.  The \textit{Fermi}/LAT data points are shown in black, with values taken from the 3FGL catalog for PKS 0637-752 and 3C 273 and the rest being upper limits taken from the analysis done for this paper.}
\end{figure*}

In Table~\ref{table:source properties} we summarize the source
properties of the four targets presented in this paper and the two
sources in which the IC/CMB model for the X-rays has already
been ruled out, PKS~0637-752 and 3C~273.  In columns 2$-$4 we report
the source redshift and corresponding angular size scale in kpc per
arcsecond and the black hole mass as $M_{BH}$, taken from the
literature where available as noted in the table. Using data from
NED, we decomposed the radio spectrum of these sources into the
extended (isotropic) ``lobe'' component and beamed ``core'' component
as in \cite{meyer2011} and shown in Figure~\ref{fig:radio seds}.  
In this Figure we plot the lobe emission with cyan squares and the core emission with blue circles (data filtered out from the fitting is shown as open circles).  The lobe and core emission were modeled with power-law and log-parabolic spectra respectively, shown as the dotted lines, with the combined fit shown as a red line.  In column 5 we give the kinetic jet power
$L_{kin}$, where we have scaled from the 300~MHz radio luminosity of
the lobes using the scaling of \cite{cavagnolo2010}.  In column 6 we
give the radio core dominance $R_{CE}$, where this is the logarithmic ratio of the
core to extended (i.e., lobe) spectrum at 1.4~GHz, and in column 7 we
give the radio crossing frequency $\nu_{cross}$, where this is the
frequency that the core spectrum crosses with the lobe spectrum.  Also following the methods described in \cite{meyer2011}, we made a simple
phenomenological fit to the synchrotron spectrum in order to
determine an approximate peak luminosity and frequency ($L_{peak}$ and
$\nu_{peak}$ respectively), which are given in columns 8 and 9 with the corresponding SED model fits shown in Figure~\ref{fig:source seds}.  
For the complete SEDs, we plot the lobe spectrum as cyan squares, \textit{Fermi}/LAT data in black, the rest of the data with blue circles (data filtered out from the fitting is shown as gray circles), and the model fits as solid red lines.  Finally we give the projected size of the source and equipartition
magnetic field strength for the brightest X-ray knot, $B_{eq}$, in columns 10 and 11.  It should be
noted that all of these sources have quite high kinetic jet powers \citep[see Figure~4 from][]{meyer2011} and black hole masses \citep[see Figure~1
from ][]{chiaberge11} compared to other radio-loud AGN generally. 

\def\arraystretch{0.8}%
\setlength\tabcolsep{3pt}
\begin{deluxetable*}{lccccccccccc}[t]
\tablecaption{\label{table:source properties} Source Properties}
\tablecolumns{11}
\tablewidth{0pt}
\tablehead{
  (1) & (2) & (3) & (4) & (5) & (6) & (7) & (8) & (9) & (10) & (11) \\
Source&z&kpc/$''$&log $M_{BH}/M_{\odot}$&log $L_{kin}$&$R_{CE}$&log $\nu _{cross}$&log $\nu L_{\nu, peak}$&log $\nu _{peak}$&Size&$B_{eq}$\\
Name&&&&$(erg\ s^{-1})$&&(Hz)&$(erg\ s^{-1})$&(Hz)&(kpc)&($10^{-4}\ G$)}
\startdata
0637-752&0.650&6.75&9.41\tablenotemark{a}&45.9&0.19&9.01&46.3&13.4&164& 1.15\\
1136-135&0.560&6.26&8.45\tablenotemark{b}&45.9&-0.72&10.7&45.4&14.1&92.9&0.470\\
1229-021 &1.05&7.91&8.70\tablenotemark{b}&46.0&0.14&9.00&46.1&14.1&149&3.20\\
1354+195 &0.720&7.03&9.44\tablenotemark{a}&45.8&-0.50&9.50&46.0&13.6&305&2.07\\
2209+080 &0.480&5.80&...&45.5&-1.5&10.4&45.6&13.4&64.4&1.00\\
3C 273&0.160&2.65&8.91\tablenotemark{c}&45.5&0.37&8.82&45.9&14.4&63.6&1.00\\
\enddata
\tablecomments{Size of the source is the projected size and determined from hot spot to counter hot spot.\\  
	a, b, c - Black hole mass taken from \cite{chen2015}, \cite{liu2006}, and \cite{kim2015} respectively.}
\end{deluxetable*}

In this paper, we aim to test whether the high level of steady
gamma-ray emission predicted under the IC/CMB model for the X-rays is
seen.  It is therefore essential that the precise level of the
expected gamma-ray emission is known.  As discussed in
\cite{georganopoulos2006}, the X-ray to gamma-ray IC component has the
same spectral shape as the radio-optical synchrotron spectrum, and the
requirement to match the observed X-rays thus fixes the rest of the
spectrum which peaks in the gamma-rays. It is thus critical that we
compile the best possible multi-wavelength SEDs for the knots, from
radio to X-rays.  We describe here the archival and new data used to
compile these SEDs for our four sources.

\subsection{VLA}
As shown in Table~\ref{table:vla}, we used the deepest available
archival \textit{VLA} imaging of sufficient quality in bands L, C, X,
and U (A or B configuration) in order to obtain a high-resolution
image with the knots distinctly resolved.  All sources were analyzed
using \texttt{CASA} version \texttt{4.7.0}. In all cases, either
3C~286 or 3C~48 was used as the flux density calibrator, and the
source itself was used for gain calibration. We applied several rounds
of self-calibration before generating the final image with
\texttt{clean}. In Table~\ref{table:vla} we list the source, the band
and configuration, the central frequency of the observation, the
project code, date of observation, time on source (TOS) in minutes,
the RMS of the final self-calibrated image in $\mu$Jy, and the size of
the restoring beam in arcseconds.

\begin{deluxetable*}{llcllccc}[t]
\tablecaption{\label{table:vla} VLA Archival Data}
\tablecolumns{8}
\tablewidth{0pt}
\tablehead{
Source & Band (Config.) & Frequency & Project & Date & TOS  & RMS       & Beam \\
       &                &    (GHz)   &         &      &  (m) & ($\mu$Jy) & ($''$) 
}
\startdata
1136-135   & C (A) & 4.860 & AH938  & 2007 Jun 23 & 59 & 154 & 0.51$\times$0.43\\
           & X (B) & 8.460 & AC689  & 2008 Nov 03 & 447& 111 & 1.01$\times$0.74\\
           & K (AB)& 22.00 & AC461  & 2002 May 27 & 107& 123 & 0.25$\times$0.15\\\rule{0pt}{4ex}
1229-021   & L (A) & 1.505 & AK95   & 1983 Oct 29 & 37 & 510 &  1.85$\times$1.26\\
           & C (A) & 4.848 & AK95   & 1983 Oct 29 & 18 & 305 &  0.61$\times$0.40 \\
           & X (A) & 8.350 & AK353  & 1994 Mar 20 & 27 & 58  & 0.28$\times$0.27\\
           & U (B) & 14.94 & AK180  & 1987 Dec 12 & 114& 72  &  0.59$\times$0.42\\\rule{0pt}{4ex}
  1354+195 & L (A) & 1.425 & AB920  & 1999 Jul 19 & 5  & 900 & 1.49$\times$1.36 \\
           & C (B) & 4.860 & AB331B & 1985 Apr 20 & 44 & 258 & 1.21$\times$1.12 \\
           & X (B) & 8.415 & BL3    & 1993 Mar 08 & 33 & 135 & 1.21$\times$0.77 \\\rule{0pt}{4ex}
  2209+080 & C (A) & 4.86  & AM723	& 2003 Aug 30 & 87 & 107 & 0.39$\times$0.36 \\
           & U (B) & 14.93 & AM723	& 2002 Aug 20 & 96 & 90  & 0.44$\times$0.41 \\
\enddata
\end{deluxetable*}


\subsection{VLBI}
We searched for archival Very Long Baseline Interferometry (VLBI)
observations of our targets which can be used to measure the pc-scale
jet proper motions. Three of the targets have no suitable archival
data or were not observed with VLBI over multiple epochs. Only one
target, PKS~1354+195, was observed four times (in 1997, 1999, 2002, and
2003) as part of the Monitoring Of Jets in Active galactic nuclei with
VLBA Experiments (MOJAVE) monitoring project \citep{lister2009,
  lister2016}, although no previous proper motion measurements were
reported. We downloaded the fully-reduced \texttt{fits} images from
the MOJAVE website, and used the publicly available Wavelet Image
Segmentation and Evaluation (WISE) code \citep{mertens2015_wise,
  mertens2016} to analyze the images for proper motions.

The WISE code comprises three main components. First, detection of
structural information is performed using the segmented wavelet
decomposition method. This algorithm provides a structural
representation of astronomical images with good sensitivity for
identifying compact and marginally resolved features and delivers a
set of two-dimensional significant structural patterns (SSP), which
are identified at each scale of the wavelet decomposition. Tracking of
these SSP detected in multiple-epoch images is performed with a
multi-scale cross-correlation algorithm. It combines structural
information on different scales of the wavelet decomposition and
provides a robust and reliable cross-identification of related SSP.

The images of PKS~1354+195 were taken with the Very Long Baseline
Array (VLBA) at 2 cm (15 GHz) on 18 August 1997, 19 July 1999, 12
August 2002, and 06 January 2003. The images were centered on the core
position and required no additional adjustment before analysis with
WISE. The final reported proper motion values were obtained using a
scale factor of 6, though results were similar across a range of
reasonable scale factors for the size of the knots in the jet. We only
report those features detected in at least 3 of four epochs.

\subsection{ALMA}
In Table~\ref{table:alma} we list the \textit{ALMA} data used in our analysis, where we give the observing band, cycle of the observations, the effective frequency of the flux measurement, the time on source,  the RMS of the final image and the size of the restoring beam.  \textit{ALMA} observations were made of the first three targets as part of Cycle 3 and 4 programs
2015.1.00932.S and 2016.1.01481.S. Both programs included observations
at band 3 and 6, however only about 40\% of the C-rated cycle 3
program was completed, and not all (A-rated) cycle 4 observations have
been released.  Forthcoming observations may improve on those
presented in this paper. PKS~1136-135 was observed in band 3 by
\textit{ALMA} as a calibrator for a project unrelated to jets
(program 2016.1.01250.S, PI: C. Peroux), and this data was kindly made
available to us by the PI. For all sources, the data was first reduced
using the provided \texttt{scriptForPI.py} script, which produces a
calibrated measurement set appropriate for imaging with
\texttt{clean}. The source scans were split off and several rounds of
phase-only (non-cumulative) self-calibration were applied before a
final (cumulative) round of amplitude and phase
self-calibration. \texttt{clean} was used in \texttt{mfs} mode, with
\texttt{nterms=1} and briggs weighting with \texttt{robust=0.5}. The
final RMS measurements correspond to the primary-beam-corrected image,
though we use the uncorrected images in the figures. In the case of
2209+080, we combined together and imaged the somewhat
lower-resolution cycle 4 data with the cycle 3 data, after core
subtraction.

\begin{deluxetable}{lcccccc}[t]
\tablecaption{\label{table:alma} ALMA Observations}
\tablecolumns{7}
\tablewidth{0pt}
\tablehead{
Source   & Band & Cycle & Freq. & TOS  & RMS       & Beam \\
         &      &       &    (GHz)   & (s)  & ($\mu$Jy) & ($''$) 
}
\startdata
1136-135 &  3   &   4   &  94.8 &  240 & 28 & 1.26$\times$1.14$''$ \\\rule{0pt}{4ex}
1229-021 &  3   &   4   &  97.5 &  66  & 171  &  0.76$\times$0.53$''$ \\
         &  6   &   4   &  233  &  131 & 93   &  0.57$\times$0.46$''$ \\\rule{0pt}{4ex}
1354+195 &  3   &   4   &  97.5 &  72  & 224  &  0.76$\times$0.53$''$ \\
         &  6   &   4   &  233  &  151 & 96   &  0.83$\times$0.56$''$ \\\rule{0pt}{4ex}
2209+080 &  3   &   3,4 &  97.5 &  375 & 246 & 0.52$\times$0.35$''$ \\
         &  6   &   4   &  233  &  454 & 161   &  0.61$\times$0.46$''$ \\
\enddata
\end{deluxetable}

\subsection{HST}
Archival \textit{HST} flux densities for the individual knots of 1354+195 were
taken from \cite{sambruna2004}, from \cite{cara2013} for 1136-136, and
from \cite{tavecchio2007} for 1229-021.

PKS~2209+080 was observed as part of our \textit{HST} program GO-13676
(PI: Meyer) and we present the results here for the first time.  This
source was observed for one orbit with the Wide Field Camera 3 (WFC3)
in the near-infrared (IR) channel with filter F160W
($\lambda$=1.6$\mu$m), and for two orbits with the Advanced Camera for
Surveys Wide Field Camera (ACS/WFC) with filter F606W
($\lambda$=6000\AA). In both cases a dithering pattern was used to
better sample the PSF, and the raw images were stacked using a routine
similar to AstroDrizzle, but better optimized for Astrometry.  In
particular, several dozen point sources in the optical image were used
as a reference frame on which all images were aligned (final error on
the registration is less than 1 mas) -- this allowed us to precisely
align the optical and IR images, which allows for better
identification of common features in both. The final pixel scale in
both images is 25 mas.

We did not attempt a galaxy subtraction, as it has a very irregular
profile and models produced by \texttt{iraf} tasks \texttt{ellipse}
and \texttt{bmodel} left large residual errors. We measured flux densities
using contours around each knot, estimating the background by moving
the same contour to 6-8 positions at the same radius from the galaxy
center. The mean background value was subtracted from the initial flux density
measurement, and errors are taken to be $\sqrt{2}$ times the standard
deviation of the background flux densities.  The flux densities were measured using the viewer utility of the \texttt{CASA} package.

\subsection{Fermi}

\begin{figure*}[ht]
\vspace{20pt}
\begin{center}
\includegraphics[width=5.5in]{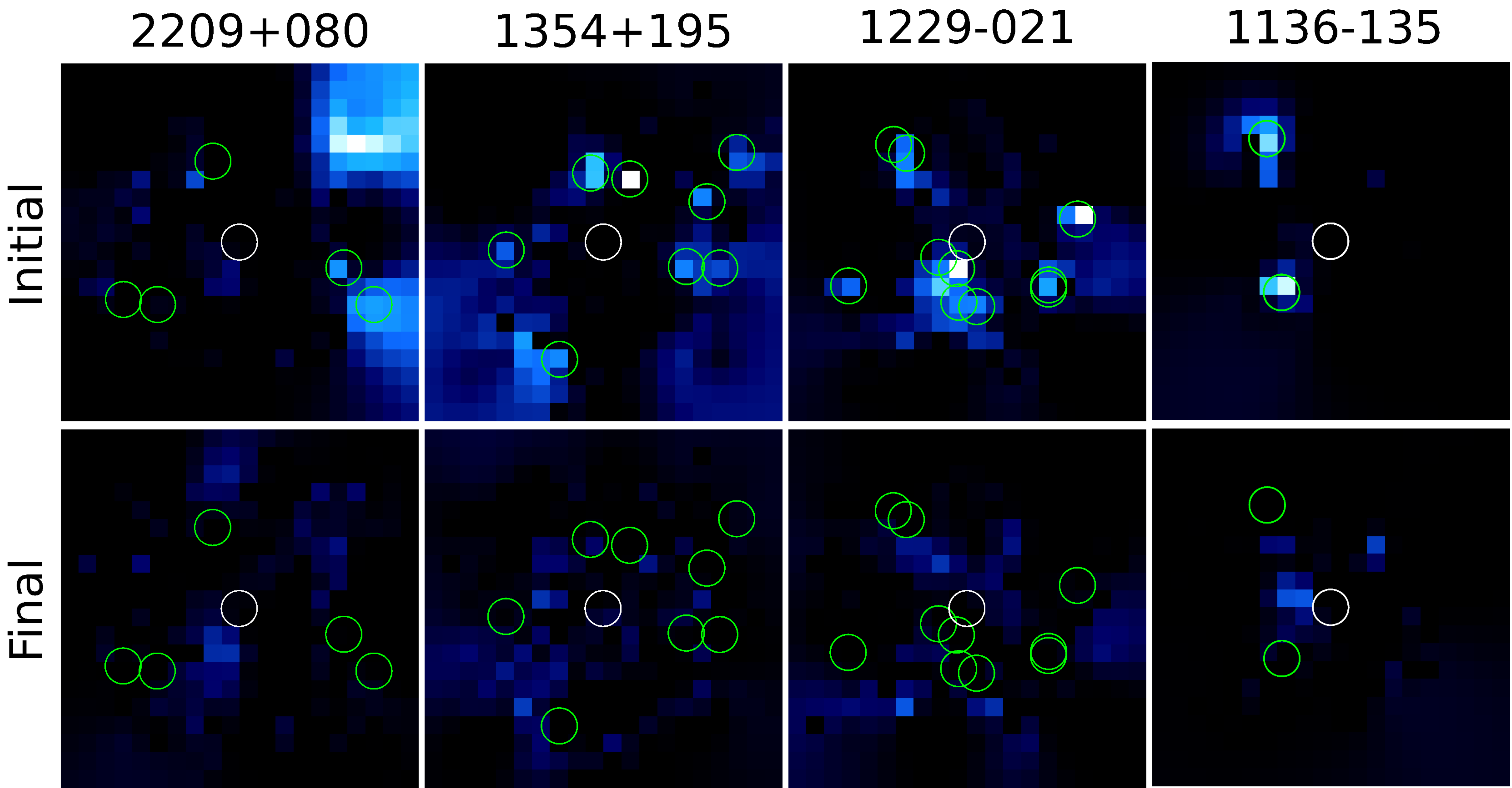}
\end{center}
\caption{\label{fig:tsmaps} Initial and final Test Statistic (TS)
  residual maps for the region around each target, with final
  localized positions of all new non-3FGL sources shown as green
  circles, and the source position shown as a white circle.  Pixels
  are one square degree.  The intensity scale ranges from a TS of 0 to
  25 in all maps and from black to white respectively.}
\end{figure*}

By design, each of the four targets was not a member of the 3FGL
4-year point source catalog for the \textit{Fermi}/LAT.  We first
checked if any of the sources would be detected using data for the
entire time range since the launch of \textit{Fermi} in 2008 (over 8
years).  \textit{Fermi}/LAT event and spacecraft data was extracted
using a $10^{\circ}$ region of interest (ROI), an energy cut of
100~MeV-100~GeV, a zenith angle cut of $90^{\circ}$, and the
recommended event class and type for point source analysis for all
sources except for PKS~1136-135 where a $7^{\circ}$ ROI was used.  The
time cuts included all available \textit{Fermi} data at the time of
analysis with corresponding mission elapse time (MET) ranges of 239557417 to 510106834 for PKS~1136-135, 239557417 to 501427132 for PKS~1229-021, 239557417 to 239557417 for PKS 1354+195, and 239557417 to 503840003 for PKS~2209+080.  Following the standard methodology for
\textit{Fermi}/LAT binned likelihood analysis, a binned counts map was
made with 30 logarithmically spaced energy bins and 0.2 degree spatial
bins in all cases.  An initial spatial and spectral model file for
each source were constructed with sources up to $10^{\circ}$ outside
the ROI using the publicly available \texttt{make3FGLxml.py} script.  This
populates the model file with point and extended sources from the
\textit{Fermi}/LAT 3FGL catalog and an extended source catalog
respectively.  Additionally, the current galactic diffuse emission
model, \texttt{gll\_iem\_v06.fits}, and recommended isotropic diffuse emission
model for point source analysis, \texttt{iso\_P8R2\_SOURCE\_V6\_v06.txt}, were
used for the analyses.  The livetime cubes were computed using 0.025
steps in $cos(\theta)$ (where $\theta$ is the inclination with respect
to the LAT's z-axis) and $1^{\circ}$ spatial binning.  Then all-sky
exposure maps were computed using the same energy binning as the
counts maps. After obtaining converged fits following the maximum
likelihood optimizations with the initial model, Test Statistic (TS)
residuals maps with $1^{\circ}$ spatial binning were created in order
to find new point sources not accounted for in the 3FGL catalog
following the procedure outlined in \cite{meyer2015}.  The TS is
defined as twice the logarithmic ratio of the maximum likelihood
calculated with an additional point source at the specified location
to the maximum likelihood without an additional source and can be
taken as roughly significance squared.  In Figure~\ref{fig:tsmaps} we show the TS residual maps created before and after adding the new sources not accounted for in the 3FGL catalog.  The new source positions are shown with green circles and the sources we obtained upper limits for this paper are shown with white circles.

In locations where TS residual values were greater than 10 we added
point sources with power-law spectra to the model files and optimized
their positions and spectral parameters.  After obtaining converged
maximum likelihood fits for the updated model file, our target source
of interest was added to the model file with a fixed photon index
(1.93 for PKS~1136-135, 2.39 for PKS~1229-021, 1.94 for PKS~1354+195, and 1.65 for PKS~2209+080).  These photon indices were derived from
the implied gamma-ray index of the IC/CMB model.  After this, the
upper limits in the five \textit{Fermi} energy bands of
100~MeV-300~MeV, 300~MeV-1~GeV, 1~GeV-3~GeV, 3~GeV-10~GeV, and
10~GeV-100~GeV were computed by running the analysis tools separately
on each data set with the appropriate energy range data cuts.

\begin{deluxetable*}{ccccccccccccc}[!htbp]
	\tablecaption{\label{table:1136 spectral data} 1136-135 Broadband Jet Knot Emission}
	\tablecolumns{13}
	\tablewidth{0pt}
	\tablehead{
		Feature & $F_{4.8GHz}$ & $F_{9.4GHz}$ & $F_{22GHz}$ & $F_{5.8\mu m}$ & $F_{3.6\mu m}$ & $F_{815nm}$ &$F_{626nm}$&$F_{555nm}$&$F_{475nm}$&$F_{1keV}$&$\alpha_{r}$ &Distance from core\\
		&(mJy)&(mJy)&(mJy)&($\mu Jy$)&($\mu Jy$)&($\mu Jy$)&($\mu Jy$)&($\mu Jy$)&($\mu Jy$)&(nJy)&&($''$)
	}
	\startdata
	$\alpha$&2.90&1.50&0.500&...&...&337&330&358&362&1.90&0.59&2.7\\
	A&3.85&2.00&0.650&...&...&237&208&212&231&1.70&0.60&4.9\\
	B&8.00&3.20&...&4.80&3.80&535&396&365&320&3.50&0.61&6.8\\
	C&30.8&9.20&...&...&...&342&...&148&...&1.80&0.80&7.8\\
	D&59.8&11.9&...&...&...&390&191&140&...&1.00&1.1&8.6\\
	E&114&26.4&...&...&...&161&...&130&...&0.700&0.97&9.3\\
	\enddata
\end{deluxetable*}

\begin{deluxetable*}{ccccccccccc}[!htbp]
	\tablecaption{\label{table:1229 spectral data} 1229-021 Broadband Jet Knot Emission}
	\tablecolumns{11}
	\tablewidth{0pt}
	\tablehead{
		Feature&$F_{1.4GHz}$&$F_{4.8GHz}$&$F_{8.4GHz}$&$F_{15GHz}$&$F_{97GHz}$&$F_{233GHz}$&$F_{692nm}$&$F_{1keV}$&$\alpha_{r}$&Distance from core\\
		&(mJy)&(mJy)&(mJy)&(mJy)&(mJy)&(mJy)&($\mu Jy$)&(nJy)&&$('')$
	}
	\startdata
	A&...&21.7&20.0&10.4&...&...&...&...&0.66&0.70\\
	BCD&274&100&65.4&42.2&10.4&4.50&0.470&8.50&0.80&1.9\\
	\enddata
\end{deluxetable*}

\begin{deluxetable*}{cccccccccccc}[!htbp]
\tablecaption{\label{table:1354 spectral data} 1354+195 Broadband Jet Knot Emission}
\tablecolumns{11}
\tablewidth{0pt}
\tablehead{
Feature&$F_{1.4GHz}$&$F_{4.8GHz}$&$F_{8.4GHz}$&$F_{97GHz}$&$F_{233GHz}$&$F_{814nm}$&$F_{586nm}$&$F_{475nm}$&$F_{1keV}$&$\alpha_{r}$&Distance from core\\
&(mJy)&(mJy)&(mJy)&(mJy)&(mJy)&($\mu Jy$)&($\mu Jy$)&($\mu Jy$)&(nJy)&&$('')$
}
\startdata
A&131&49.2&32.0&5.20&3.50&0.530&0.300&0.195&16.1\tablenotemark{$\dagger$}&0.77&1.7\\
B&30.8&14.2&7.60&1.55&0.477&0.0770&0.0400&0.0470&0.700&0.79&4.0\\
\enddata
\tablecomments{\tablenotemark{$\dagger$}{This flux density is in dispute, and more appropriately may be a non-detection with an upper limit of 0.39~nJy (private communication with D. Schwartz)}}
\end{deluxetable*}

\begin{deluxetable*}{ccccccccc}[!htbp]
	\tablecaption{\label{table:2209 spectral data} 2209+080 Broadband Jet Knot Emission}
	\tablecolumns{9}
	\tablewidth{0pt}
	\tablehead{
		Feature & $F_{4.8GHz}$ & $F_{15GHz}$ & $F_{233GHz}$ & $F_{1.6\mu m}$ & $F_{600nm}$ & $F_{1keV}$ & $\alpha_{r}$ &Distance from core\\
		&(mJy)&(mJy)&(mJy)&($\mu Jy$)&($\mu Jy$)&(nJy)&&($''$)
	}
	\startdata
	A&30.0&15.1&2.22&4.40&2.41&...&0.78&0.52\\
	B&21.0&12.2&2.18&1.64&0.570&...&0.60&1.3\\
	C&24.0&13.1&2.11&0.950&0.260&...&0.64&2.0\\
	D&8.20&5.00&1.15&0.170&...&...&0.52&3.2\\
	E&29.7&16.4&2.02&1.58&0.430&4.76&0.71&4.7\\
	\enddata
\end{deluxetable*}

\section{Results}
\label{sec_results}
\begin{figure*}[!htbp]
\vspace{20pt}
\begin{center}
\includegraphics[width=4.5in]{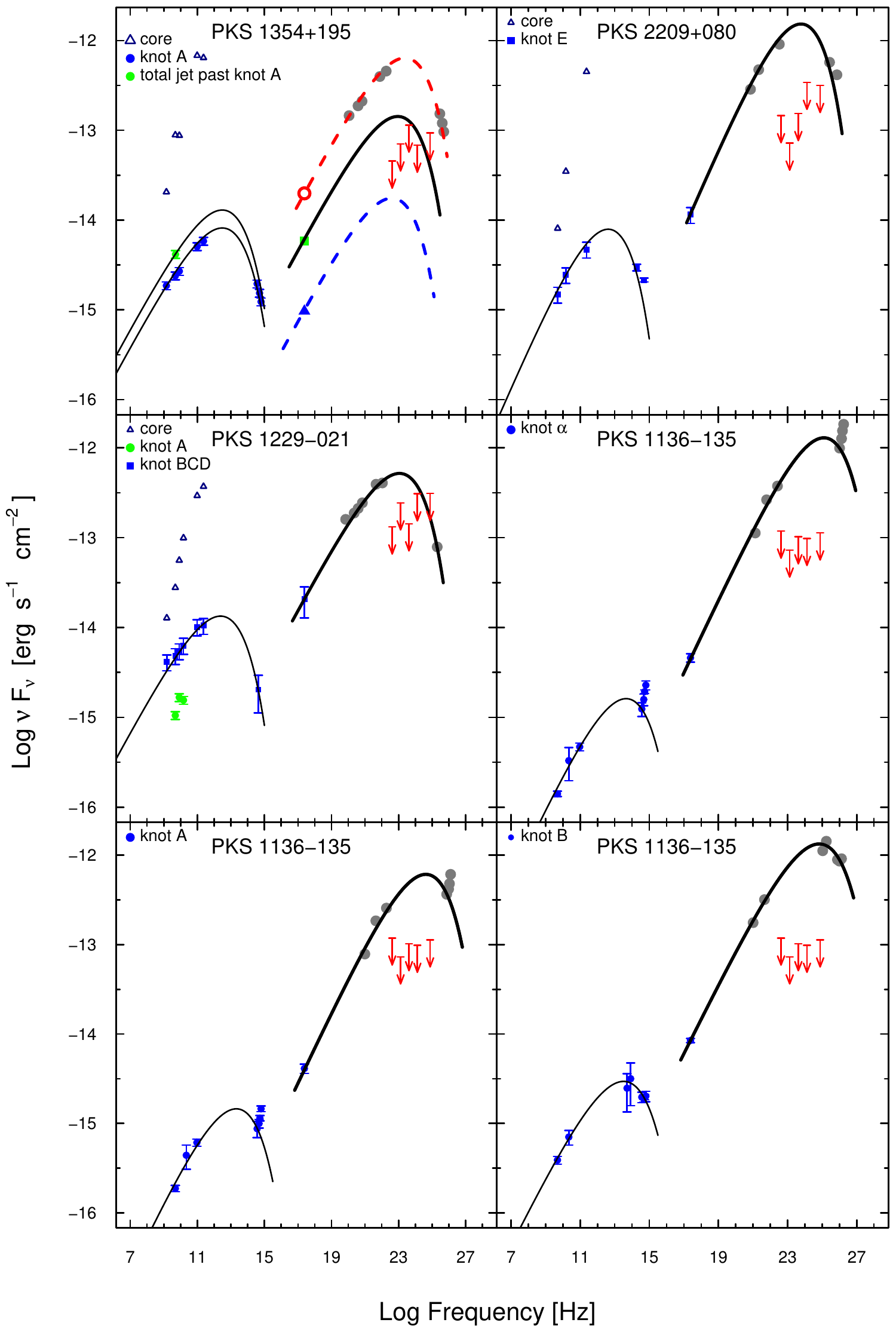}
\end{center}

\caption{\label{fig:jet_SEDs} SEDs for the jet knots of
  PKS 1354+195 (top left), PKS 2209+080 (top right), PKS 1229-021
  (middle-left), and PKS 1136-135 (middle right and bottom row).  Thin black curves show phenomenological
  synchrotron model fits and the thick black curves show the
  corresponding IC/CMB model curves normalized to match the X-ray flux densities.
  The gray points represent where the radio through optical data
  points would lie on the IC/CMB model curves after shifting in
  frequency and luminosity by the appropriate amount.  The 95\%
  \textit{Fermi} upper limits are shown in red.  For
  PKS 1229-021, blue data points correspond to the combined data for
  knots B,C, and D.  The green data points for PKS~1354+195 correspond to the combined data for all knots past knot A with the X-ray point being a lower limit and shown as a square (all data for PKS~1354+195 except the \textit{Fermi} limits and the radio/\textit{ALMA} data of knot A are taken from \cite{sambruna2004}).  The red open circle and red dashed line for PKS~1354+195 shows the original 8.2 nJy X-ray flux density for knot A reported by \cite{sambruna02} with the corresponding IC/CMB model fit.  The blue triangle for PKS~1354+195 shows the 0.39 nJy upper limit for the X-ray flux density of knot A obtained in private communication with D. Schwartz, with the corresponding IC/CMB model curve shown as a blue dashed line.  The thick black line in the PKS~1354+195 SED is the IC/CMB model curve for the total jet past knot A.
 The X-ray and optical data for PKS 1229-021 are taken from
  \cite{tavecchio2007}; the X-ray data for PKS 2209+080 are taken from
  \cite{jorstad2006}, and the X-ray, infrared, and optical data for
  PKS 1136-135 are taken from \cite{cara2013}.}
\end{figure*}

In Tables~\ref{table:1136 spectral data}-\ref{table:2209 spectral data} we present the flux densities from radio through X-ray wavelengths of each identified knot of sources 1136-135, 1229-021, 1354+195 and 2209+080, respectively, with the corresponding radio
spectral index and distance of the knot from the core.  The radio
spectral index, $\alpha_{r}$, is defined using the convention
$F_{\nu}\propto \nu^{-\alpha_{r}}$.  We show in
Figure~\ref{fig:jet_SEDs} the broadband SEDs for (typically) the most
X-ray prominent knots in each source. In the case of PKS~1354+195, we
also show the total X-ray flux density of the jet after knot~A, and in PKS~1229-021, we
show the combined flux density of knots B, C, and D which could not be
separately resolved in all imaging.

%

In Figure~\ref{fig:jet_SEDs}, observed flux densities are plotted (in $\nu F_{\nu}$) as blue and green
squares with thin solid black lines showing the synchrotron model fits
to the radio-optical data. In all cases the radio-optical spectrum is
modeled as a power law with scaled exponential cutoff as described in
the appendix.  The thicker black lines are the IC/CMB model fits to
the X-ray flux densities.  The IC/CMB model curves are copies of the synchrotron
curves shifted in frequency and luminosity by an amount proportional
to $\delta/B$ and $(\delta/B)^{2}$ respectively, where $\delta/B$ is
the only free parameter in the shift set by the requirement to
reproduce the observed X-ray flux density and the equations governing the
shift are given in the appendix \citep[see also][]{georganopoulos2006}\footnote{$B$ is the magnetic field strength of
the emitting region and $\delta$ is the Doppler factor given by
$\delta=\Gamma/(1-\beta cos\theta)$.  Here $\Gamma$ is the bulk
Lorentz factor, $\beta$ is the bulk speed flow scaled by c, and
$\theta$ is the jet angle to the line-of-sight.}.  The 95\%
\textit{Fermi} upper limits are shown as red arrows.  As can be seen,
the \textit{Fermi} upper limits are well below the IC/CMB model curves
in each SED and we can reject the IC/CMB model as the X-ray emission
mechanism for each source.

The results of the \textit{Fermi} analyses, corresponding to the
plotted limits, are given in Table~\ref{table:Fermi Results}, where
for all 5 \textit{Fermi} bands we give the minimum and maximum energy
in columns 4 and 5, the logarithmic mean frequency of the band in
column 6 corresponding to the frequency of the predicted IC/CMB
flux densities (column 7) and the observed 95\% upper limits (column 8). In columns 9 and
10 we give the confidence with which we can rule out the IC/CMB model
for each X-ray knot as a percentage and sigma level. Finally, we
derive and list in column 11 the upper limit on the Doppler factor
assuming an equipartition magnetic field.  Combining the individual
band results using the inverse normal method, we find the IC/CMB model
is ruled out at the $14.3\sigma$, $6.1\sigma$, $4.7\sigma$, and $>15.2\sigma$-levels for PKS~1136-135, PKS~1229-021, PKS~1354+195, and PKS~2209+080 respectively. Additionally, in Figure~\ref{fig:jet_SEDs} we plot in gray closed
circles where the observed synchrotron data points would lie in the
shifted IC/CMB spectrum.  This was done to emphasize the point that
the \textit{Fermi} upper limits directly violate the IC/CMB predicted
flux densities from observed portions of the synchrotron spectrum.

\setlength\tabcolsep{4pt}
\def\arraystretch{0.7}%
\begin{deluxetable*}{lcccccccccc}[!htbp]
\tablecaption{\label{table:Fermi Results} Results of the Fermi Data Analysis}
\tablecolumns{11}
\tablewidth{0pt}
\tablehead{(1) & (2) & (3) & (4) & (5) & (6) & (7) & (8) & (9) & (10) & (11) \\
Source&Feature&Band&$E_{1}$&$E_{2}$&log&Predicted&95\% &\%&$\sigma$ & $\delta$\\
Name&&&&&freq&$\nu F_{\nu,IC/CMB}$&$\nu F_{\nu}$ limit&Ruled Out&&limit\\
&&&(GeV)&(GeV)&(Hz)&($erg\ s^{-1} cm^{-2}$)&($erg\ s^{-1} cm^{-2}$)&&&}
\startdata
1136-135&B&1&0.1&0.3&22.6&$5.7\times 10^{-13}$&$1.18\times 10^{-13}$&99.996&3.94&5.4\\
&&2&0.3&1&23.1&$7.7\times 10^{-13}$&$7.27\times 10^{-14}$&>99.99999&6.29&\\
&&3&1&3&23.6&$9.9\times 10^{-13}$&$1.02\times 10^{-13}$&>99.99999&7.58&\\
&&4&3&10&24.1&$1.2\times 10^{-12}$&$9.81\times 10^{-14}$&>99.99999&8.13&\\
&&5&10&100&24.7&$1.3\times 10^{-12}$&$1.13\times 10^{-13}$&>99.99999&6.01&\\
1229-021&BCD\tablenotemark{$\dagger\dagger$}&1&0.1&0.3&22.6&$4.9\times 10^{-13}$&$1.32\times 10^{-13}$&99.97&3.40&8.0\\
&&2&0.3&1&23.1&$5.2\times 10^{-13}$&$2.44\times 10^{-13}$&99.91&3.11&\\
&&3&1&3&23.6&$4.8\times 10^{-13}$&$1.43\times 10^{-13}$&99.998&4.07&\\
&&4&3&10&24.1&$3.9\times 10^{-13}$&$3.07\times 10^{-13}$&97.6&1.98&\\
&&5&10&100&24.7&$2.1\times 10^{-13}$&$3.10\times 10^{-13}$&85.8&1.07&\\
1354+195&jet\tablenotemark{$\dagger$}&1&0.1&0.3&22.6&$1.4\times 10^{-13}$&$4.53\times 10^{-14}$&99.83&2.92&6.0\\
&&2&0.3&1&23.1&$1.4\times 10^{-13}$&$7.05\times 10^{-14}$&99.27&2.44&\\
&&3&1&3&23.6&$1.3\times 10^{-13}$&$1.14\times 10^{-13}$&96.60&1.83&\\
&&4&3&10&24.1&$9.9\times 10^{-14}$&$6.83\times 10^{-14}$&97.98&2.05&\\
&&5&10&100&24.7&$5.1\times 10^{-14}$&$9.36\times 10^{-14}$&88.49&1.20&\\
2209+080&E&1&0.1&0.3&22.6&$1.1\times 10^{-12}$&$1.45\times 10^{-13}$&>99.99999&5.23&8.6\\
&&2&0.3&1&23.1&$1.3\times 10^{-12}$&$7.19\times 10^{-14}$&>99.99999&>8.2&\\
&&3&1&3&23.6&$1.5\times 10^{-12}$&$1.54\times 10^{-13}$&>99.99999&>8.2&\\
&&4&3&10&24.1&$1.5\times 10^{-12}$&$3.42\times 10^{-13}$&>99.99999&7.87&\\
&&5&10&100&24.7&$1.1\times 10^{-12}$&$3.15\times 10^{-13}$&99.999&4.34&\\
\enddata
\tablecomments{\tablenotemark{$\dagger$}{Data was combined for the whole jet past knot A}}
\tablenotemark{$\dagger\dagger$}{Knot BCD includes the combined data from knots B, C, and D.}
\end{deluxetable*}

\begin{figure*}[ht]
	\vspace{20pt}
	\begin{center}
		\includegraphics[width=6in]{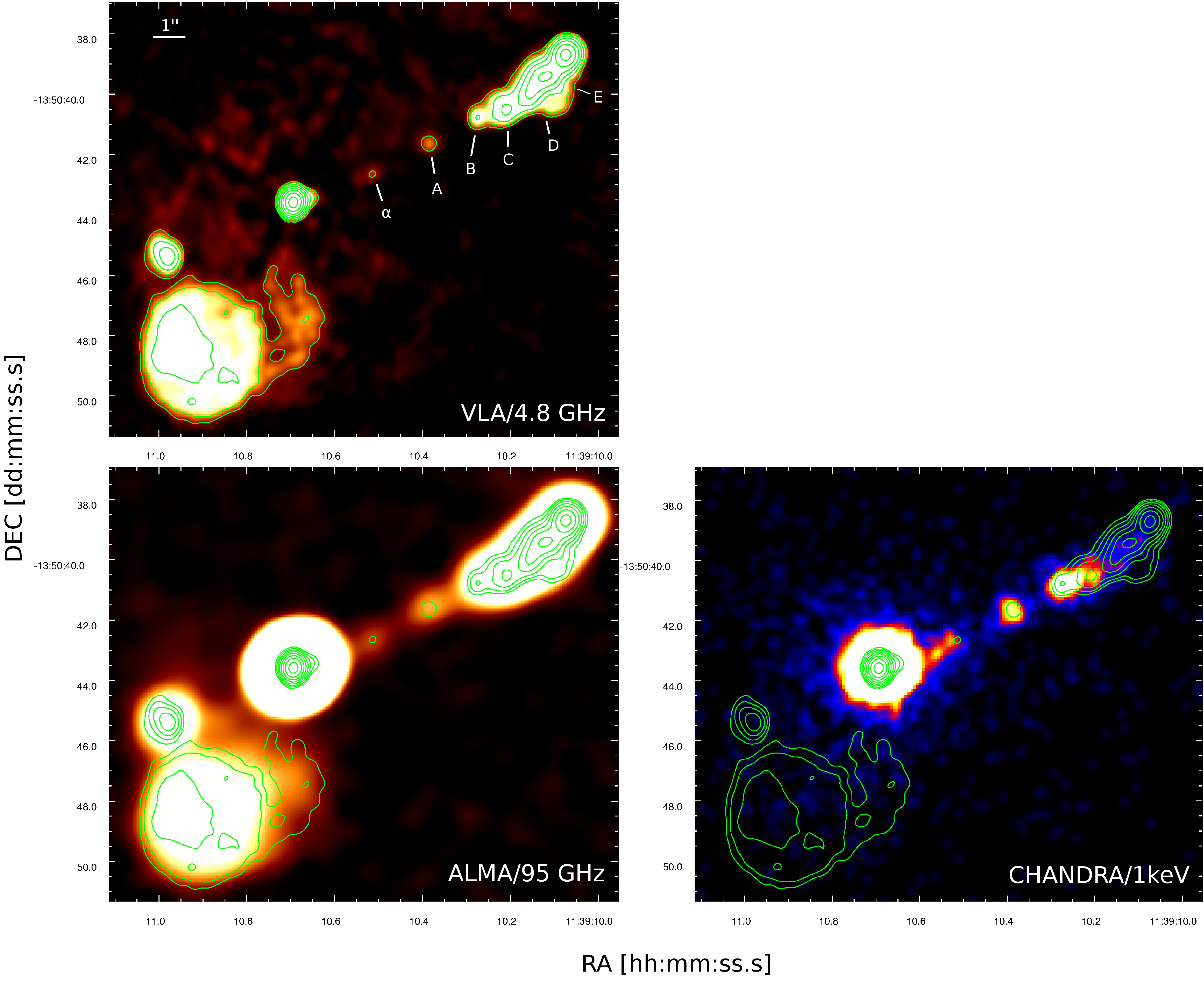}
	\end{center}
	\caption{\label{fig:1136 images} Multi-wavelength images of
		PKS~1136-135 where 4.8~GHz \textit{VLA} contours were used for all
		images.  Contours are spaced by a factor of 2, with a base level of
		1.5 mJy.  The top left panel is a 4.8~GHz \textit{VLA} image, the
		bottom left panel is \textit{ALMA} band 3, and the bottom right
		panel is a 1~keV \textit{Chandra} image.  The \textit{Chandra} image
		is smoothed with a Gaussian with a radius of 3 pixels.}
\end{figure*}

\begin{figure*}[ht]
	\vspace{20pt}
	\begin{center}
		\includegraphics[width=5in]{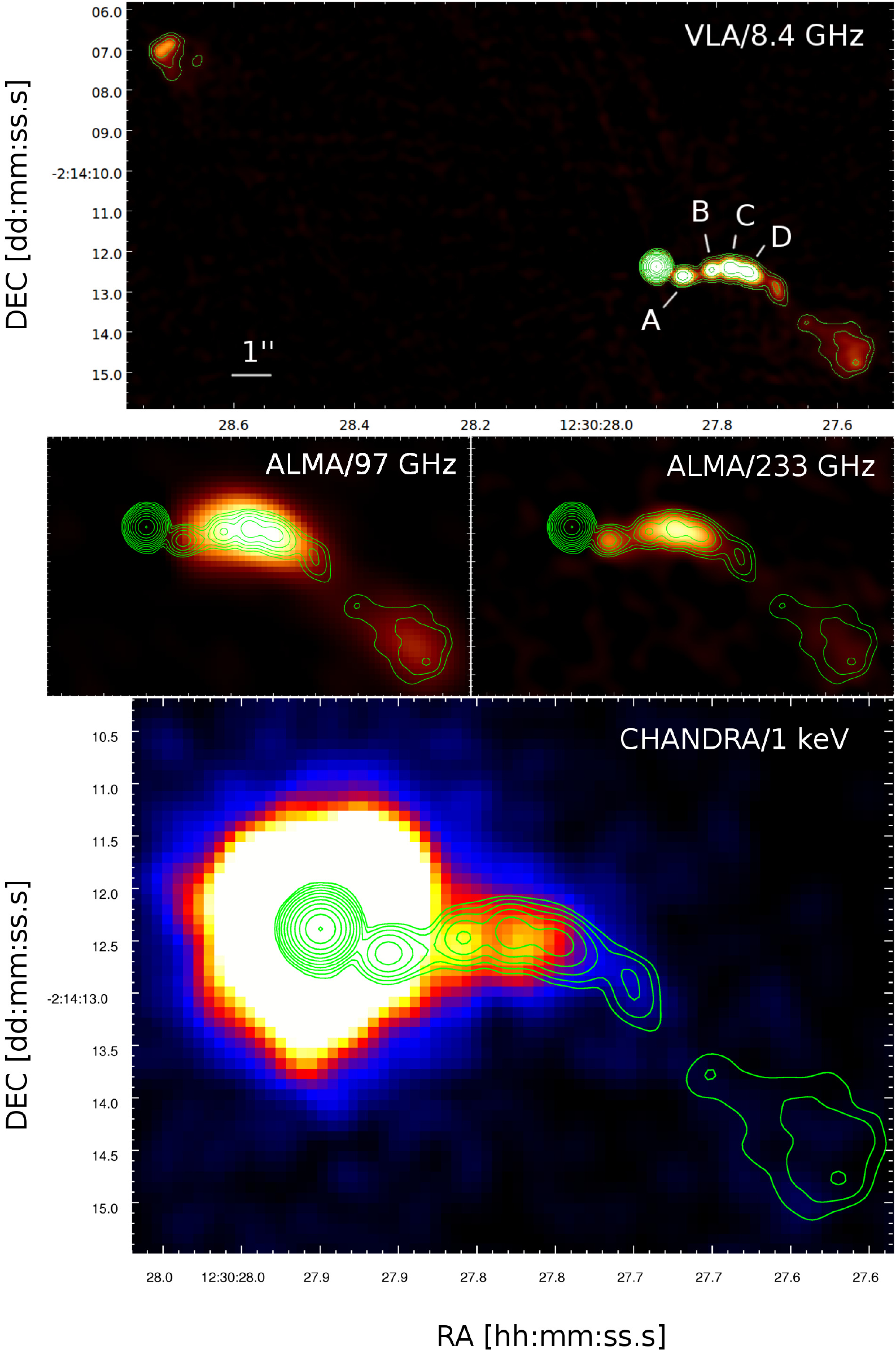}
	\end{center}
	\caption{\label{fig:1229 images} Multi-wavelength images for
		PKS~1229-021 where 8.4 GHz \textit{VLA} contours were used for all
		images.  The contours are spaced by factors of 2, with a base level
		of 0.63 mJy.  The upper panel is an 8.4~GHz \textit{VLA} image, the
		middle panels are \textit{ALMA} band 3 and 6 images, and the lower
		panel is a \textit{Chandra} 1~keV image.  The \textit{Chandra} image
		is smoothed with a Gaussian with a radius of 3 pixels.  The
		\textit{ALMA} images are focused on the jet and hot spot, and are
		core-subtracted.}
\end{figure*}

Radio, sub-mm, optical (for 2209+080) and X-ray images of each source in this study are shown in
Figures~\ref{fig:1136 images}, \ref{fig:1229 images}, \ref{fig:1354
	images}, and ~\ref{fig:2209 images} with the knot identifications
used in this paper (which in some cases vary from previous usage in the literature). In
all four cases there is no detectable counter-jet at any wavelengths,
but we do detect the counter-hot spots in the radio and in some cases
with \textit{ALMA}.  We now discuss the observations of each source in
turn.

\begin{figure*}[ht]
	\vspace{20pt}
	\begin{center}
		\includegraphics[width=4.5in]{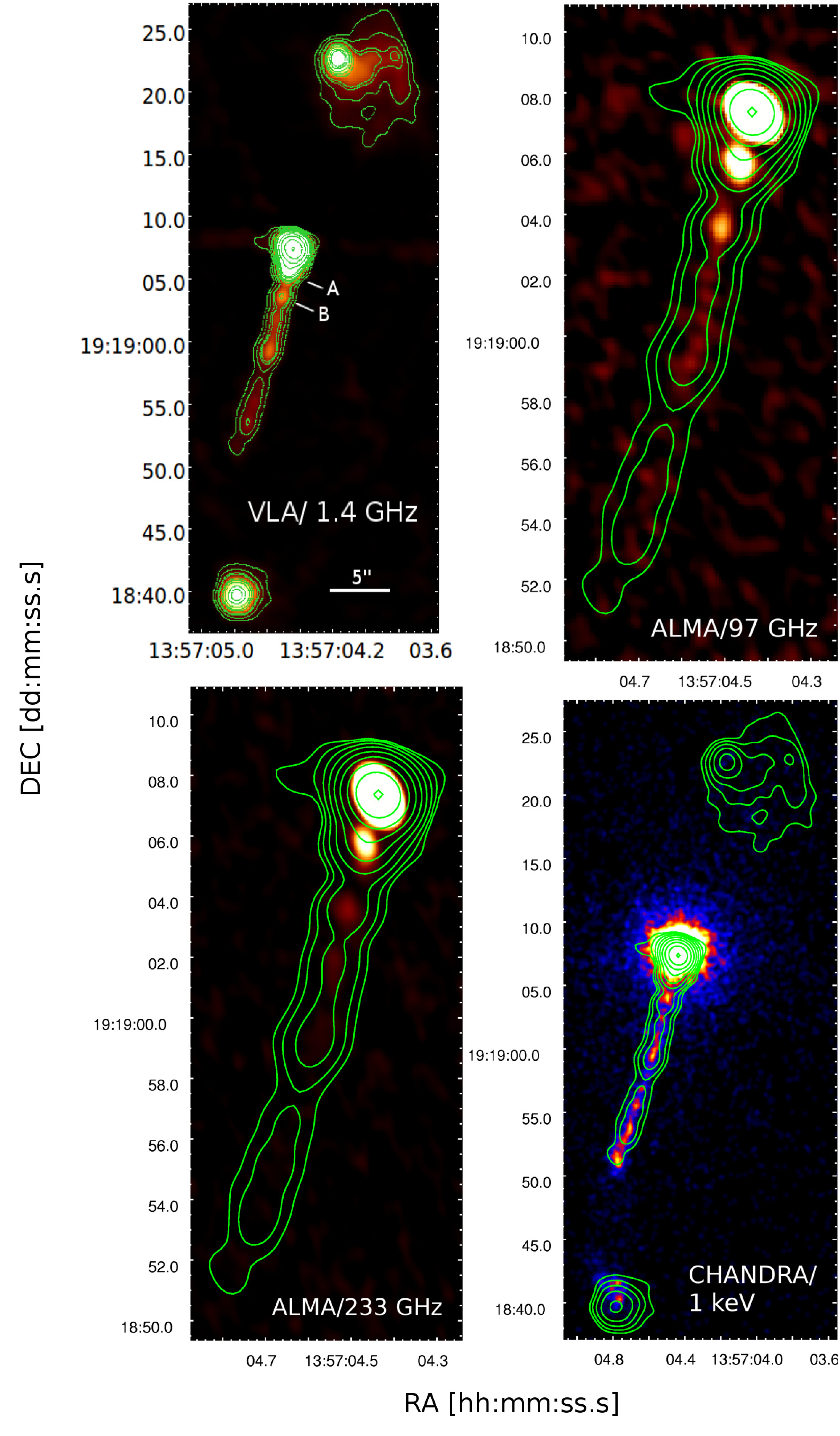}
	\end{center}
	\caption{\label{fig:1354 images} Multi-wavelength images of
		PKS~1354+195 where 1.4~GHz \textit{VLA} contours were used for all
		images.  The contours are spaced by factors of 2, with a base level
		of 5.2 mJy.  The upper left panel is a 1.4~GHz \textit{VLA} image,
		the upper right panel is \textit{ALMA} band 3, the lower left panel
		is \textit{ALMA} band 6, and the lower right panel is a 1~keV
		\textit{Chandra} image.  \textit{ALMA} images are focused on the jet
		to show the knot structure.  The \textit{Chandra} image is smoothed
		with a Gaussian with a radius of 3 pixels.}
\end{figure*}

\begin{figure*}[ht]
\vspace{20pt}\begin{center}
\includegraphics[width=6in]{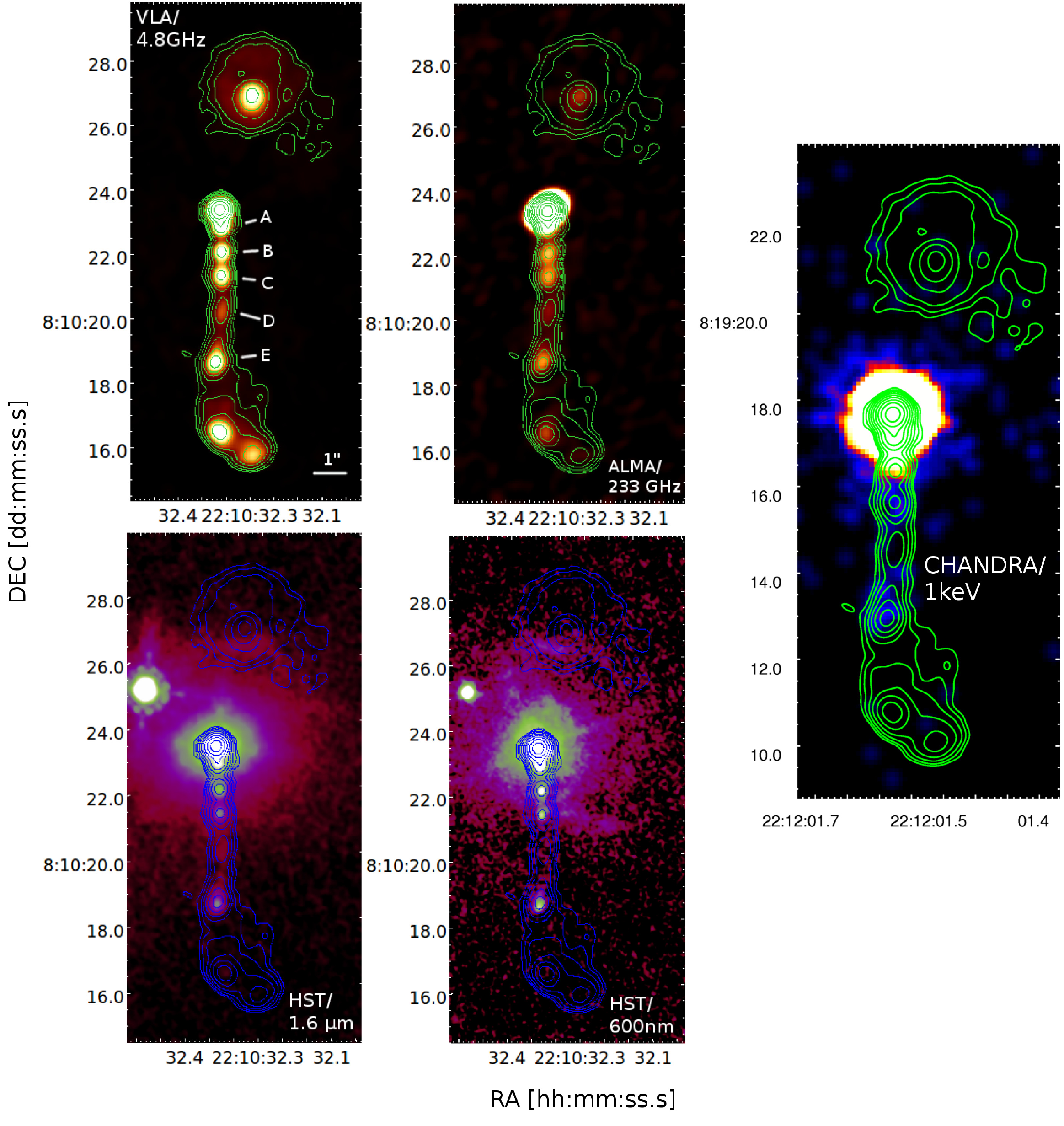}
\end{center}
\caption{\label{fig:2209 images} Multi-wavelength images of
  PKS~2209+080 with 4.8~GHz \textit{VLA} contours for all images.  The
  contours are spaced by factors of 2, with a base level of 56~mJy.
  The upper left panel is a 4.8~GHz \textit{VLA} image, the upper
  middle panel is an \textit{ALMA} band 6 image, the right panel is a
  1~keV \textit{Chandra} image, and the lower left images are
  \textit{HST} 1.6$\mu$m and 600~nm.  The 600nm image is smoothed with
  a Gaussian with a radius of 3 pixels.  }
\end{figure*}

\subsection{PKS 1136-135}

PKS 1136-135 is a powerful FR~II source previously observed by \cite{sambruna06} with \textit{Chandra} and \textit{HST}, where they modeled the X-ray emission with the IC/CMB model.  Our multi-wavelength images are shown in Figure~\ref{fig:1136 images}, where we see a straight and knotty radio jet, with knots brighter towards the hot spot in the radio.  Knots B, C, D, and E are not separately resolved by \textit{ALMA} so we do not include \textit{ALMA} flux densities for the jet SED.  We show in Figure~\ref{fig:jet_SEDs} the SED of knots $\alpha$, A, and B, where our gamma-ray upper limits are well below the extrapolated IC/CMB emission in all three cases.  Our limits were not deep enough to rule out IC/CMB for the last 3 knots which had low IC/CMB predicted gamma-ray flux densities.  In knots $\alpha$ and A there is an upturn in the SED at 815~nm which
lies above a single power-law fit for the synchrotron spectrum.
Fitting these UV-upturns with an IC/CMB model is problematic since
this would require very high levels of X-ray emission not observed by
\textit{Chandra}.  Therefore, these UV-upturns imply a second higher
energy electron distribution also responsible for the X-ray emission, similar to that seen in 3C 273.  \cite{cara2013} showed with \textit{HST} polarimetry that all of the jet knots had fractional polarization measures in excess of 30\% (except knots $\alpha$ and B where they found $2\sigma$ upper limits of 15\% and 14\% respectively).  For the cases where the UV emission is clearly part of the second spectral component, these high degrees of polarization provide strong evidence against the IC/CMB model.  However, our gamma-ray upper limits provide another independent line of evidence ruling out the IC/CMB model in knot A, while our upper limits show that the IC/CMB model is ruled out for knots $\alpha$ and B which did not have high degrees of polarization.  Fitting the UV data with an IC/CMB model for knots $\alpha$ and A would also predict much higher X-ray flux densities than observed, though the UV spectrum is also much harder than would be consistent with an IC/CMB model fit.

\subsection{PKS~1229-021}

PKS~1229-021 was previously observed and modeled as an IC/CMB X-ray
jet by \cite{tavecchio2007}, where we use their reported \textit{HST}
and \textit{Chandra} densities in our SED (Figure~\ref{fig:jet_SEDs}).
The radio images shown in Figure~\ref{fig:1229 images} show four
well-defined 'cannonball'-like knots downstream of the
core. Interestingly, similar periodic knot structures are seen in all
six of the jets discussed in this paper.  As shown in
Figure~\ref{fig:1229 images}, knots B, C, and D are not separately
resolved by \textit{ALMA} and \textit{Chandra}; thus we combined the
data for these knots, which are labeled as knot BCD.  Past this
combined feature, we see the jet is considerably bent, implying a
change of jet direction.  Our \textit{ALMA} images (which are
core-subtracted) are focused on the jet in order to emphasize the knot
structure; we found the western hot spot was only detected by
\textit{ALMA} in band 3 and not in band 6.  The \textit{HST} data
presented in \cite{tavecchio2007} shows that only the combined knot
BCD was detected in the \textit{HST} imaging while knot A is not
detected. With only one optical data point, the synchrotron SEDs for
both of these features are not as well-constrained as for our other
sources.  Therefore we believe this source would be a good candidate
for follow-up \textit{HST} observations in order to further constrain
the SED.

\subsection{PKS~1354+195}
This source is shown in Figure~\ref{fig:1354 images}, where the radio
observations show a straight jet with detailed hot
spot structure for the northern hot spot. The \textit{VLA} and
\textit{ALMA} observations again show a well-defined cannonball-like
knot structure for knots A and B.  This jet was previously observed and modeled as an IC/CMB X-ray
jet by \cite{sambruna02} and \cite{sambruna2004}, where we use their
reported \textit{HST} and \textit{Chandra} flux densities from the 2004 publication. Initially we focused on knot A, which was reported as brightest in the X-rays.  This knot is approximately
2$''$ from the core and initially reported in \cite{sambruna02} to
have a flux density of 8.2 nJy (corresponding to about 135 counts) and later
revised to 16.1 nJy in \cite{sambruna2004}.  However, we have recently
learned (D. Schwartz, private communication) that more recent, deeper
\emph{Chandra} observations of PKS~1354+195 suggest that knot A is in
fact not detectable separate from the bright quasar core, with an
upper limit of $<$0.39~nJy, after taking into account careful modeling
of the PSF from the core. We plot the 8.2 nJy \textit{Chandra} flux density in
Figure~\ref{fig:jet_SEDs} as a red open circle. While this flux density implies that the IC/CMB model is ruled out at a very high level of
significance (see red dashed line), the much lower revised flux density limit
(blue triangle and dashed line) does not. In light of this revision,
we also examined the newer \emph{Chandra} observations and can confirm
that we do not find a significant excess of counts at the position of
knot~A in any of the later observations, suggesting that the apparent
excess in the early observation (Chandra Observation ID \#2104) was a
statistical fluctuation.  The jet of PKS~1354+195 is quite straight up
to just before the hotspot, suggesting no major bends in or out of our
line-of-sight.  If knot A has the same radio to X-ray ratio as knot B,
we would expect an X-ray flux density of about 2.4 nJy, however the very bright
core makes it difficult to conclude much from knot A.  Turning to the
remaining knots in the jet, and assuming \emph{some} level of X-ray
flux density from knot A, we produce the \emph{lower limit} X-ray flux density shown
as a green square in the SED figure with a value of 2.4 nJy. Note this is the approximately the same X-ray flux density calculated for knot A assuming the same radio to X-ray ratio as knot B.  Using this and the spectral shape
determined by knots A and B together from radio-optical, we see that
the IC/CMB model is still ruled out, though the case is not as strong
as in the other sources presented in this paper.

Note that in Table~\ref{table:Fermi Results}, the predicted IC/CMB
level, the 95\% upper limits, and the corresponding confidence and
limits on $\delta$ all correspond to the more reliable estimates
derived from the knots excluding knot~A.


\subsubsection{VLBI proper motions}
\begin{figure}[ht]
\vspace{20pt}
\begin{center}
\includegraphics[width=3.5in]{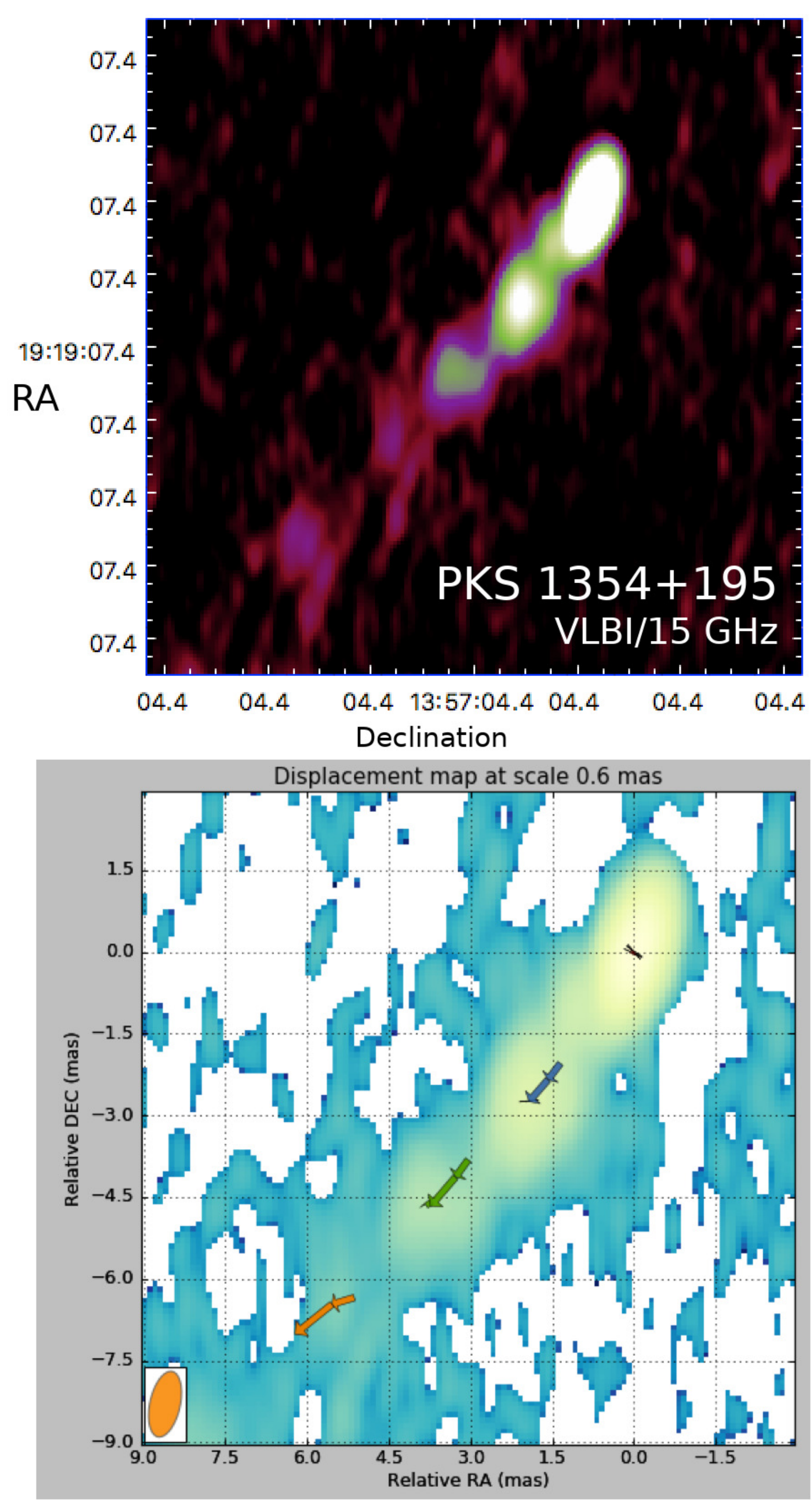}
\end{center}
\caption{\label{fig:1354_vlbi} At top: the parsec-scale jet of PKS~1354+195 in 2002 taken with VLBA at 15~GHz \citep{lister2009}. Below: the results of the WISE proper motion analysis of the four available VLBA epochs, with arrows showing the motions from epoch-to-epoch for three components in the jet (the core is stationary and consistent with no motion as expected). }
\end{figure}

PKS~1354+195 was the only source for which archival VLBA data were
available for a proper motion analysis. The results of the WISE
analysis reveal three moving features in the jet. We show the pc-scale
jet in the 2002 epoch and the results of the WISE feature tracking in
Figure~\ref{fig:1354_vlbi}. Ordered by distance from the core, the
moving features show angular motions of 0.17$\pm$0.01, 0.21$\pm$0.01,
and 0.26$\pm$0.20 mas/year.  At the scale of 7.32 parsecs per mas,
this corresponds to a apparent proper motions of 4.00$\pm$0.24$c$,
4.95$\pm$0.24$c$ and 6.1$\pm$4.7$c$.  The last of these components is
only detected in the last three images and has a rather large error
bar, but the other two components are secure detections in all four
epochs.  We re-analyzed the data using a variety of possible scales
for the components, and found results very consistent with the above.

\subsection{PKS~2209+080}   

PKS~2209+080 was previously observed and modeled as an IC/CMB X-ray
jet by \cite{jorstad2006}, where we used their X-ray flux density in our SED.
As shown in Figure~\ref{fig:2209 images}, PKS~2209+080 has a very
knotty and straight jet with the southern hot spot showing two
resolved components.  Our \textit{ALMA} and \textit{HST} observations
resolve all of the identified radio knots.  As can be seen in both
\textit{HST} images, the host galaxy has an irregular tail structure
suggestive of a possible recent merger.  Only the southern hot spot is
detected in the infrared as shown in the lower-middle panel of
Figure~\ref{fig:2209 images}.  The only knot reported to show X-ray
emission by \cite{jorstad2006} was knot E.  In Figure~\ref{fig:2209
  knot seds} we show the SEDs of knots A through D which are upstream of knot
E with data shown as squares and synchrotron model fits as solid
lines.  In knots A through C there is an upturn in the SED at 600~nm which
lies above a single power-law fit for the synchrotron spectrum.  Similar to the cases of PKS~1136-135 and 3C~273, fitting these UV-upturns with an IC/CMB model would require X-ray flux densities much higher than observed by \textit{Chandra}.

\begin{figure*}[ht]
\vspace{20pt}
\begin{center}
\includegraphics[width=6in]{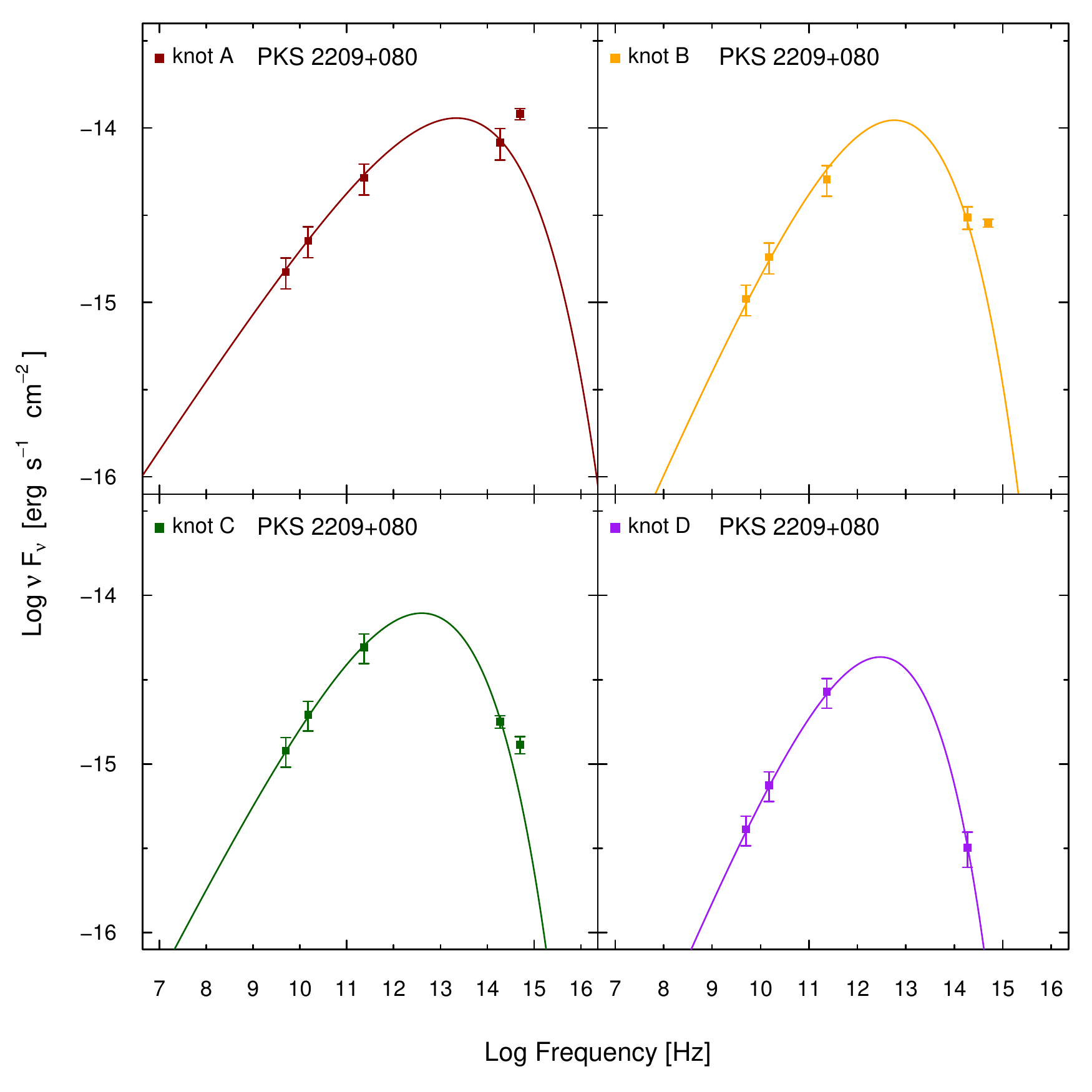}
\end{center}
\caption{\label{fig:2209 knot seds}SEDs of knots A through D for PKS~2209+080. Single power-law fits with scaled exponential cutoffs are used to fit the radio data and first optical data point for each knot.}
\end{figure*}

\section{Discussion}
\label{sec_disc}

\subsection{IC/CMB now ruled out for six sources}

The IC/CMB model for the bright and/or hard large scale
jet X-ray emission associated with powerful quasars has now been ruled
out for six sources on the basis of over-predicting the observed
gamma-ray flux.
For each source presented in this
paper, we found upper limits to the gamma-ray flux well below the levels
predicted by the IC/CMB model using observations from the
\textit{Fermi}/LAT.  For the sources not presented in this paper,
3C~273 and PKS~0637-752, previous \textit{Fermi}/LAT observations
showed similar IC/CMB violating upper limits (\citealt{meyer2014};
\citealt{meyer2015}; \citealt{meyer17}).  However, over-predicting the
observed gamma-ray flux is not the only line of evidence against the
IC/CMB model for these sources.

\subsection{UV upturns}

In the case of PKS~1136-135, a high degree of UV polarization with
fractional polarization measures in excess of 30\% for several knots
was detected in the rising second component of the jet SEDs
\citep{cara2013}.  This implies a synchrotron origin for the second
spectral component of these knots (see \citealt{jester2006};
\citealt{atoyan2004}; \citealt{harris2004}; \citealt{hardcastle2006};
\citealt{uchiyama2006}; \citealt{kataoka06}) as the IC/CMB emission is
expected to have very low polarization (see \citealt{McNamara2009};
\citealt{uchiyama2008}) while synchrotron radiation can be highly
polarized.  These UV-upturns in the knot SEDs are clearly seen in the
case of knots A-C for PKS~2209+080 shown in Figure~\ref{fig:2209 knot
  seds} and knots $\alpha$ and A for PKS~1136-135 shown in
Figure~\ref{fig:jet_SEDs}.  The optical spectrum for these knots is
consistently harder than can be fit with a single power law from radio
to optical since the exponential cutoff occurs well before the
optical.  In these cases, fitting the UV-component of the SED with an
IC/CMB model would predict much higher X-ray flux densities than are observed.
Furthermore, knots $\alpha$ and A in PKS~1136-135 have a UV spectrum
which is much harder than an IC/CMB model fit.  These inconsistencies
are additional lines of evidence against the IC/CMB model and imply
another electron distribution higher in energy to produce the
UV-upturns and the X-ray emission.

\subsection{Misalignment of the non-IC/CMB sources}

\begin{figure*}[ht]
\vspace{20pt}
\begin{center}
\includegraphics[width=6in]{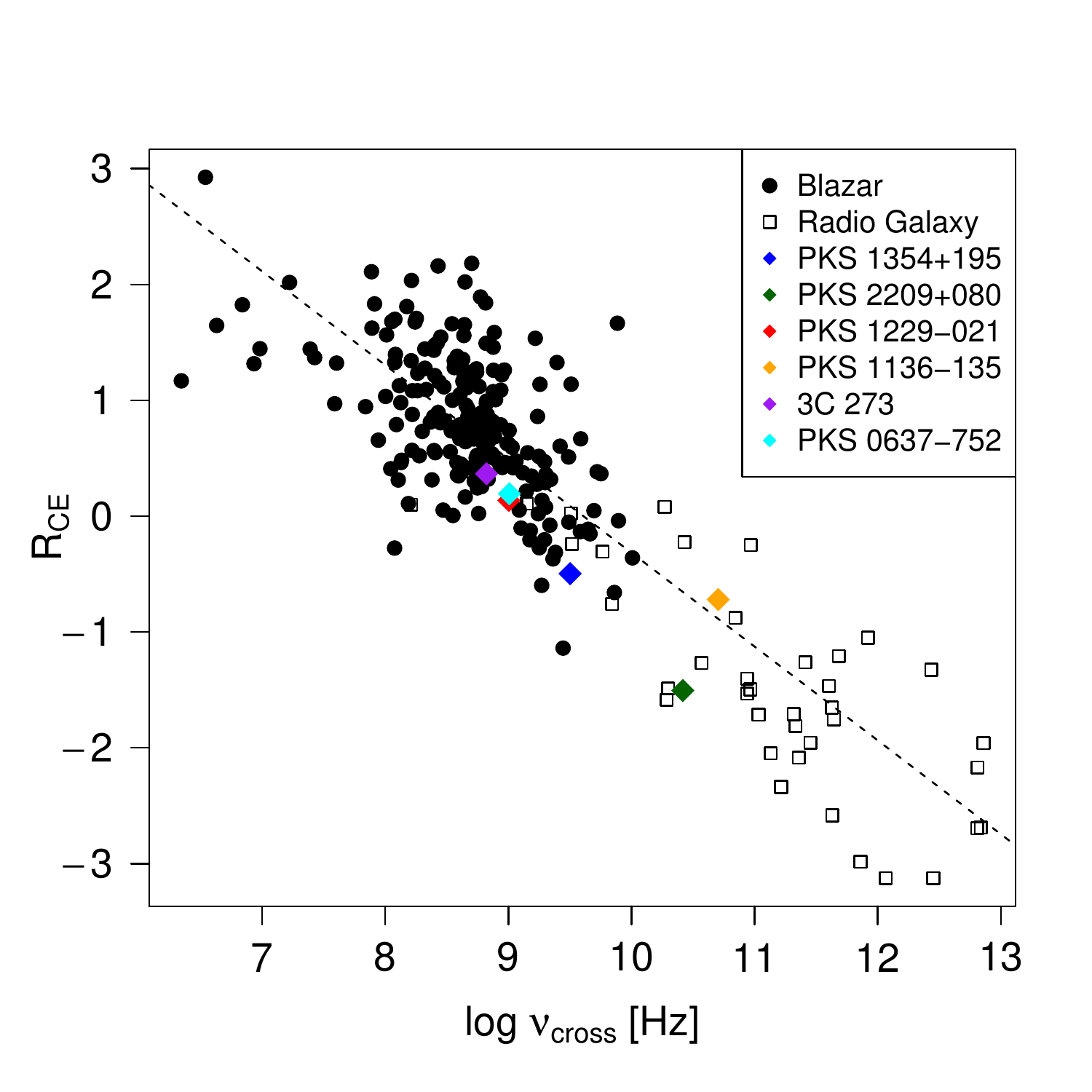}
\end{center}
\caption{\label{fig:core dominance}Plot adapted from \cite{meyer2011}.  The radio core dominance, $R_{CE}$, plotted against $\nu_{cross}$, the frequency at which the jet becomes dominant.  Radio  galaxies  are  shown  as empty squares and blazars as  filled  circles.  The broken line shown is the linear correlation between the plotted variables (r=0.87).  Plotted in diamonds are the six sources where we have now ruled out the IC/CMB model.}  
\end{figure*}

Also in tension with the IC/CMB model is the apparent misalignment of the sources.  The IC/CMB model is able to reproduce the observed bright X-ray flux densities by requiring very small jet angles to the line of sight so the X-ray emission is highly relativistically beamed (\citealt{tavecchio2000}; \citealt{celotti2001}).  It has been previously shown by \cite{meyer2011} that blazars have a higher radio core dominance, $R_{CE}$, and lower crossing frequency, $\nu_{cross}$, than radio galaxies.  Figure~\ref{fig:core dominance}, adapted from \cite{meyer2011}, shows the locations of all six sources in the $R_{CE}$-$\nu_{cross}$ plane in addition to other previously classified blazars and radio galaxies.  
As can be seen in Figure~\ref{fig:core dominance}, our sources do not have the low crossing frequencies and high radio core dominances expected for highly aligned blazars.  Additionally, PKS~1136-135 and PKS~2209+080 are well within the radio galaxy category suggesting they are fairly misaligned.  Also apparent from the total SEDs shown in Figure~\ref{fig:source seds} is that none of these sources have a Compton dominance greater than 1, which would be expected for highly aligned jets \citep[see Figure~3 from ][]{meyer12}.       

\subsection{X-ray versus Radio beaming patterns}

One property of the IC/CMB model is that it predicts a different
relativistic beaming pattern than synchrotron emission.  The expected beaming pattern for IC/CMB radiation is $L=L'\delta^{p+1+2\alpha_{r}}$, with p=2 for a continuous flow and p=3 for discrete moving blobs (\citealt{dermer95}; \citealp{georganopoulos01}).  L is the luminosity in the galaxy frame assuming isotropy, $L^{\prime}$ is the solid-angle integrated luminosity in the jet frame, $\alpha_{r}$ is the radio spectral index, and $\delta$ is the Doppler factor.  In the case of synchrotron emission, the expected beaming pattern is given by $L=L'\delta^{p+\alpha_{r}}$ \citep{dermer95}.  

\begin{figure*}[ht]
	\vspace{20pt}
	\begin{center}
		\includegraphics[width=6in]{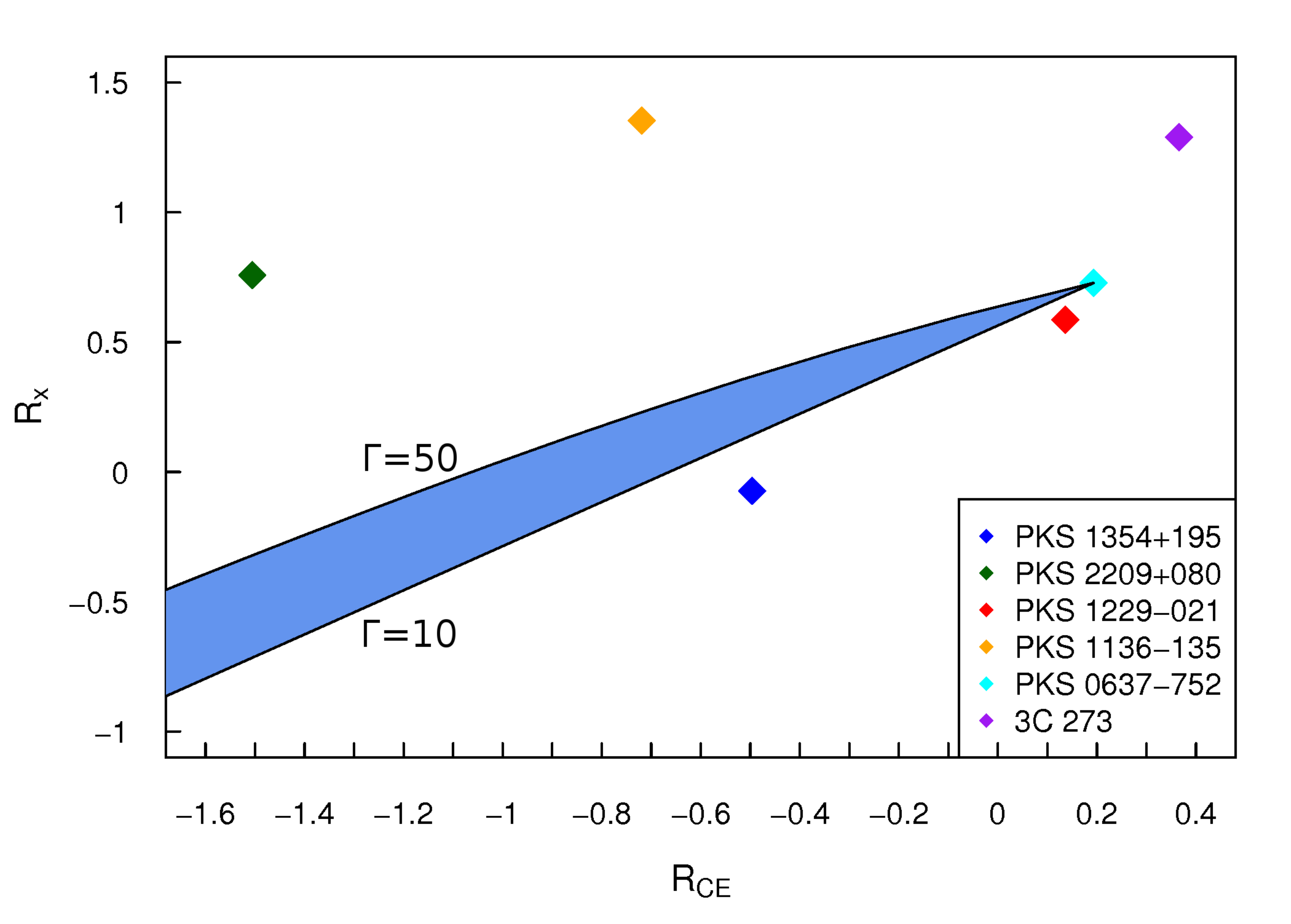}
	\end{center}
	\caption{\label{fig:xray dominance}The X-ray dominance, $R_{x}$, plotted against the core dominance, $R_{CE}$, for the six sources where the IC/CMB model has now been ruled out shown as diamonds.  Error bars are not shown as they are on the order of the size of the data points.  We plot $R_{x}$ for the brightest X-ray knot in each source.  Smaller values of $R_{CE}$ correspond to more misaligned sources.  The black lines show the expected result of misaligning the jet of PKS~0637-752 assuming an initial value of $\delta=10$ for the knot with an initial misalignment of $5.73^{\circ}$.  The upper and lower black lines show the cases where the Lorentz factor of the core is 50 and 10 respectively, with the blue shaded region showing the possible range between those values.}
\end{figure*}

Since larger angles of the
jet to the line-of-sight decrease the value of $\delta$, the observed
IC/CMB flux should fall off faster than the synchrotron flux as source misalignment increases due to the stronger dependence on $\delta$
of IC/CMB emission.  We define a quantity we refer to as the X-ray
dominance, $R_{x}$, as the logarithmic ratio of the 1~keV X-ray flux density to the 8.6~GHz radio flux density (in $\nu F_{\nu}$).  In Figure~\ref{fig:xray dominance} we show the
X-ray dominance plotted against the core dominance for the X-ray
brightest knots of the six sources in which IC/CMB is now ruled out as the X-ray emission mechanism,
where the core dominance is a measure of the relative misalignment
of the jet angle to the line-of-sight as can be seen in Figure~\ref{fig:core dominance}.  Using the above relativistic beaming equations for IC/CMB and synchrotron emission, it can be shown that $R_{x}\propto\delta_{knot}^{1+\alpha_{knot}}$ and $R_{CE}\propto\delta_{core}^{p+\alpha_{core}}$.  Using these equations, we plot the projected misalignment of PKS~0637-752 in Figure~\ref{fig:xray dominance} as solid black curves assuming $\alpha_{core}=0$, $\alpha_{knot}=0.7$, and p=2 for the core which reflects the most likely case in which the core is a standing shock and not a discrete moving feature.  The misalignment curves are plotted for a core $\Gamma$ of 10 and 50 with the shaded blue region representing the range of misalignment curves for a core $\Gamma$ between 10 and 50.  As can be seen in Figure~\ref{fig:xray dominance},  the other five sources do not fall within this misalignment zone of PKS~0637-752 as would be
expected for the stronger beaming of the IC/CMB emission, admittedly
with a small sample of sources.  The lack of correlation between $R_{x}$ and $R_{CE}$ suggests the radio-optical synchrotron emission and X-ray emission for these knots have the same relativistic beaming profile and not a different one as would be expected if the X-rays were produced by the IC/CMB mechanism.  However, the large range in X-ray dominance also suggests that there is some intrinsic variability of the X-ray component within the sample.  A larger sample should allow a more careful examination of the expected correlation, which we leave to future work.  

\subsection{Morphology and Host Properties of the non-IC/CMB sources}

All of these sources show what we term a cannon-ball jet knot structure in the radio with quasi-periodic knot spacing.  It has been suggested that this quasi-periodic spacing of the jet knots may be due to re-confinement shocks of the external medium, accretion disk instabilities, or a binary super massive black hole \citep{godfrey2012}.  Another interesting characteristic of the radio morphology is the bright elongated structure at the end of the jet and lack of extended lobe emission for the sources 3C~273 \citep{bahcall95} and PKS~1136-135 (though there is clearly lobe emission connected to the counter jet in PKS~1136-135) which may be identifiable with the "nose cone" jet structure seen in Magnetohydrodynamics simulations.  These nose cone structures are thought to occur when the jets are magnetically confined, and rather than the plasma deflecting back from the interface between the jet and intergalactic medium to form the lobes, it is collected in the end of the jet between the Mach disk and leading bow shock forming the "nose cone".  One observational signature associated with the nose cone jets is enhanced radio emission in the nose cone due to an enhanced magnetic field strength and thus synchrotron emissivity, but a decrease in the flux from higher energies due to a lack of particle acceleration in this region from shocks (see \citealt{clarke86} for a discussion of nose cone jets).  This observational characteristic appears to be true in the 2 cases of 3C~273 and PKS~1136-135 in which we can identify a possible nose cone structure.

The host galaxy of PKS~2209+080 which is visible in our \textit{HST} observations shows a clear tidal tail structure which is suggestive of the galaxy having undergone a recent merger.  The disturbed elliptical host galaxy of 3C~273 has also shown some evidence of being in a recent merger \citep{martel03}.  These findings are consistent with the work by \cite{chiaberge15} suggesting that all radio-loud AGN have undergone major recent mergers.   

\subsection{Alternatives to IC/CMB}

One emission mechanism which has not been ruled out for the
X-rays in MSC jets is synchrotron radiation from a second higher energy population
of electrons.  If these X-ray knots are due to a second
electron energy distribution, this means that these jets can exhibit
highly efficient in situ particle acceleration very far from the
central engine (on the order of kpcs).  These electrons would need to
be highly energetic (at a few to hundreds of TeV energies) and in
cases where the IR/optical flux is totally consistent with a single
spectrum from radio to optical, have a lower energy cutoff such that they do not overproduce
the observed IR/optical emission.  One consequence of these multi-TeV
energy electrons is that they should inverse-Compton scatter the CMB
to TeV gamma-rays which may be detectable with the upcoming
\textit{CTA} for low redshift sources (\citealt{meyer2015};
\citealt{georganopoulos2006}).

Alternatively, it is possible these X-rays are due to hadronic
emission processes whereby the X-rays are produced by direct proton
synchrotron radiation or from the synchrotron radiation produced by
secondary electrons resulting from photo-hadronic interactions like
the Bethe-Heitler process or photopion production (see
\citealt{petropoulou2017}; \citealt{Bhattacharyya16};
\citealt{kusunose17}).  Like synchrotron models, and unlike IC/CMB, hadronic models are highly tunable and can be made to reproduce most observed SEDs since there are a large number of parameters to adjust.  However, one distinguishing characteristic of hadronic models would be the expected neutrino emission from charged pion or neutron decay resulting from interactions between high energy protons and photons \citep{mannheim89}.  It has been claimed that the bright gamma-ray flare observed in the blazar PKS~B1424-418 is hadronic in nature due to the detection of a PeV-energy neutrino with a high degree of positional and temporal coincidence \citep{kadler} by the IceCube neutrino detector.  However, the detection rate of IceCube high energy neutrinos is so low that this is not expected to be a very robust means of distinguishing between hadronic and leptonic models for blazars.  Additionally, in the case of PKS~0637-752, hadronic
emission models require highly super-Eddington jet powers
\citep{kusunose17}.

These models for the X-ray emission do not have many of the
problematic features of the IC/CMB model. They do not require these
jets to remain highly relativistic on large scales or be highly
aligned.  The observed "knottiness" of
these X-ray jets is well-explained by the much stronger radiative
losses for these emission mechanisms.  Finally, the second synchrotron
model naturally explains the high degrees of polarization measured in
the UV-upturns for PKS~1136-135 \citep{cara2013} and 3C~273
\citep{jester07}.

\section{Summary \& Conclusions}
\label{sec_summ}
In this paper we have shown that the IC/CMB model cannot explain the X-ray emission for the large scale MSC jets of the powerful
quasars PKS~1136-135, PKS~1229-021, PKS~1354+195, and PKS~2209+080 at
the $14.3\sigma$, $6.1\sigma$, $4.7\sigma$, and  $>15.2\sigma$ levels
respectively.  The IC/CMB model was ruled out due to IC/CMB violating
gamma-ray upper limits with the \textit{Fermi}/LAT, where the
predicted inverse-Compton gamma-ray flux is fixed by the
requirement of fitting the observed X-ray flux density with the IC/CMB model.  In the case of
PKS~1136-135, knots $\alpha$ and A also show a steeply rising optical
spectrum which would predict much higher X-ray flux densities than observed if
fit with an IC/CMB model.  These sources now represent an additional
three to the list of MSC jets where the IC/CMB model has
already been conclusively ruled out: PKS 0637-752, 3C 273, and PKS
1136-135 ({PKS~1136-135 has already been ruled out on the basis of UV polarization}).  Additionally, our \textit{HST} observations for PKS
2209+080 show a newly detected optical jet with possible tidal tails in the
host galaxy. Based on the radio core dominance and crossing frequencies
of PKS~1136-135 and PKS~2209+080, we have shown that these jets
must be fairly misaligned, not extremely well-aligned blazars as is
required under the IC/CMB model.  Since these results lead us to
reject the IC/CMB model as the emission mechanism for these MSC jets, we must consider alternative emission mechanisms such as a second population of electrons producing a second synchrotron component or hadronic emission models.  In the case of the second synchrotron model, it would be important to understand how the high energy
electron distribution is created and what mechanisms might give rise
to such highly efficient particle acceleration in situ kpcs from the
central engine.
\section{Appendix}

\subsection{Synchrotron fits}

The empirical fits to the observed synchrotron SEDs are simple power laws with scaled exponential cutoffs, corresponding to a simple power-law electron energy distribution with maximum Lorentz factor.  They have the following form: 

\begin{equation}
\nu f_{\nu}=N\left(\frac{\nu}{10^{10} Hz}\right)^{\gamma}exp\left(-\left(\frac{\nu}{\nu_{1}}\right)^{\beta}\right)
\end{equation}

In this equation $\nu$ is the observed frequency of the radiation, $\gamma$ the power-law index, $\nu_{1}$ the frequency at which the exponential turnover begins, $\beta$ the steepness of the cutoff, and N is the normalization of the spectrum which has the units $erg\ cm^{-2}\ s^{-1}$.  

\subsection{IC/CMB shifting equations}

The IC/CMB model has the following form for the shifting in luminosity and frequency of the lower energy synchrotron spectrum, first derived in \cite{georganopoulos2006}:

\begin{equation}
\frac{\nu_{c}}{\nu_{s}}=\frac{\nu_{CMB}}{e(B/\delta)/\left[2\pi m_{e}c(1+z)\right]}
\end{equation}

\begin{equation}
\frac{L_{c}}{L_{s}}=\frac{32U_{CMB}\left(1+z\right)^{4}}{3(B/\delta)^{2}}
\end{equation}

B is the magnetic field strength, e is the elementary charge, $U_{CMB}$ is the CMB energy density at the current epoch, $\nu_{CMB}$ the CMB peak frequency at the current epoch, z is the redshift, $m_{e}$ is the electron mass, c is the speed of light, and $\delta$ is the Doppler factor.  Note the only free parameter in this shift is $B/\delta$ which becomes fixed upon fitting the X-ray component of the SED.  If one assumes an equipartition magnetic field, this shift is only parametrized by $\delta$.  Also note that these shifts should preserve the same spectral index for both the synchrotron and inverse-Compton components, although this is not always seen for these MSC jets.    


\bibliography{bibliography}

\begin{thebibliography}{}
\expandafter\ifx\csname natexlab\endcsname\relax\def\natexlab#1{#1}\fi

\bibitem[{{Aharonian}(2002)}]{aharonian2002}
{Aharonian}, F.~A. 2002, \mnras, 332, 215

\bibitem[{{Arshakian} \& {Longair}(2004)}]{arshakian2004}
{Arshakian}, T.~G., \& {Longair}, M.~S. 2004, \mnras, 351, 727

\bibitem[{{Atoyan} \& {Dermer}(2004)}]{atoyan2004}
{Atoyan}, A., \& {Dermer}, C.~D. 2004, \apj, 613, 151

\bibitem[{{Bahcall} {et~al.}(1995){Bahcall}, {Kirhakos}, {Schneider}, {Davis},
  {Muxlow}, {Garrington}, {Conway}, \& {Unwin}}]{bahcall95}
{Bahcall}, J.~N., {Kirhakos}, S., {Schneider}, D.~P., {et~al.} 1995, \apjl,
  452, L91

\bibitem[{{Bhattacharyya} \& {Gupta}(2016)}]{Bhattacharyya16}
{Bhattacharyya}, W., \& {Gupta}, N. 2016, \apj, 817, 121

\bibitem[{{Cara} {et~al.}(2013){Cara}, {Perlman}, {Uchiyama}, {Cheung},
  {Coppi}, {Georganopoulos}, {Worrall}, {Birkinshaw}, {Sparks}, {Marshall},
  {Stawarz}, {Begelman}, {O'Dea}, \& {Baum}}]{cara2013}
{Cara}, M., {Perlman}, E.~S., {Uchiyama}, Y., {et~al.} 2013, \apj, 773, 186

\bibitem[{{Cavagnolo} {et~al.}(2010){Cavagnolo}, {McNamara}, {Nulsen},
  {Carilli}, {Jones}, \& {B{\^i}rzan}}]{cavagnolo2010}
{Cavagnolo}, K.~W., {McNamara}, B.~R., {Nulsen}, P.~E.~J., {et~al.} 2010, \apj,
  720, 1066

\bibitem[{{Celotti} {et~al.}(2001){Celotti}, {Ghisellini}, \&
  {Chiaberge}}]{celotti2001}
{Celotti}, A., {Ghisellini}, G., \& {Chiaberge}, M. 2001, \mnras, 321, L1

\bibitem[{{Chartas} {et~al.}(2000){Chartas}, {Worrall}, {Birkinshaw},
  {Cresitello-Dittmar}, {Cui}, {Ghosh}, {Harris}, {Hooper}, {Jauncey}, {Kim},
  {Lovell}, {Mathur}, {Schwartz}, {Tingay}, {Virani}, \&
  {Wilkes}}]{chartas2000}
{Chartas}, G., {Worrall}, D.~M., {Birkinshaw}, M., {et~al.} 2000, \apj, 542,
  655

\bibitem[{{Chen} {et~al.}(2015){Chen}, {Zhang}, {Zhang}, \& {Yu}}]{chen2015}
{Chen}, Y.~Y., {Zhang}, X., {Zhang}, H.~J., \& {Yu}, X.~L. 2015, \mnras, 451,
  4193

\bibitem[{{Chiaberge} {et~al.}(2015){Chiaberge}, {Gilli}, {Lotz}, \&
  {Norman}}]{chiaberge15}
{Chiaberge}, M., {Gilli}, R., {Lotz}, J.~M., \& {Norman}, C. 2015, \apj, 806,
  147

\bibitem[{{Chiaberge} \& {Marconi}(2011)}]{chiaberge11}
{Chiaberge}, M., \& {Marconi}, A. 2011, \mnras, 416, 917

\bibitem[{{Clarke} {et~al.}(1986){Clarke}, {Norman}, \& {Burns}}]{clarke86}
{Clarke}, D.~A., {Norman}, M.~L., \& {Burns}, J.~O. 1986, \apjl, 311, L63

\bibitem[{{Clautice} {et~al.}(2016){Clautice}, {Perlman}, {Georganopoulos},
  {Lister}, {Tombesi}, {Cara}, {Marshall}, {Hogan}, \&
  {Kazanas}}]{clautice2016}
{Clautice}, D., {Perlman}, E.~S., {Georganopoulos}, M., {et~al.} 2016, \apj,
  826, 109

\bibitem[{{Dermer}(1995)}]{dermer95}
{Dermer}, C.~D. 1995, \apjl, 446, L63

\bibitem[{{Dermer} \& {Atoyan}(2004)}]{dermer2004}
{Dermer}, C.~D., \& {Atoyan}, A. 2004, \apjl, 611, L9

\bibitem[{{Fanaroff} \& {Riley}(1974)}]{fanaroffriley}
{Fanaroff}, B.~L., \& {Riley}, J.~M. 1974, \mnras, 167, 31P

\bibitem[{{Georganopoulos} {et~al.}(2001){Georganopoulos}, {Kirk}, \&
  {Mastichiadis}}]{georganopoulos01}
{Georganopoulos}, M., {Kirk}, J.~G., \& {Mastichiadis}, A. 2001, \apj, 561, 111

\bibitem[{{Georganopoulos} {et~al.}(2006){Georganopoulos}, {Perlman},
  {Kazanas}, \& {McEnery}}]{georganopoulos2006}
{Georganopoulos}, M., {Perlman}, E.~S., {Kazanas}, D., \& {McEnery}, J. 2006,
  \apjl, 653, L5

\bibitem[{{Godfrey} {et~al.}(2012){Godfrey}, {Lovell}, {Burke-Spolaor},
  {Ekers}, {Bicknell}, {Birkinshaw}, {Worrall}, {Jauncey}, {Schwartz},
  {Marshall}, {Gelbord}, {Perlman}, \& {Georganopoulos}}]{godfrey2012}
{Godfrey}, L.~E.~H., {Lovell}, J.~E.~J., {Burke-Spolaor}, S., {et~al.} 2012,
  \apjl, 758, L27

\bibitem[{{Hardcastle}(2006)}]{hardcastle2006}
{Hardcastle}, M.~J. 2006, \mnras, 366, 1465

\bibitem[{{Hardcastle} {et~al.}(2016){Hardcastle}, {Lenc}, {Birkinshaw},
  {Croston}, {Goodger}, {Marshall}, {Perlman}, {Siemiginowska}, {Stawarz}, \&
  {Worrall}}]{Hardcastle2016}
{Hardcastle}, M.~J., {Lenc}, E., {Birkinshaw}, M., {et~al.} 2016, \mnras, 455,
  3526

\bibitem[{{Harris} \& {Krawczynski}(2006)}]{harris2006}
{Harris}, D.~E., \& {Krawczynski}, H. 2006, \araa, 44, 463

\bibitem[{{Harris} {et~al.}(2004){Harris}, {Mossman}, \& {Walker}}]{harris2004}
{Harris}, D.~E., {Mossman}, A.~E., \& {Walker}, R.~C. 2004, \apj, 615, 161

\bibitem[{{Jester} {et~al.}(2006){Jester}, {Harris}, {Marshall}, \&
  {Meisenheimer}}]{jester2006}
{Jester}, S., {Harris}, D.~E., {Marshall}, H.~L., \& {Meisenheimer}, K. 2006,
  \apj, 648, 900

\bibitem[{{Jester} {et~al.}(2007){Jester}, {Meisenheimer}, {Martel}, {Perlman},
  \& {Sparks}}]{jester07}
{Jester}, S., {Meisenheimer}, K., {Martel}, A.~R., {Perlman}, E.~S., \&
  {Sparks}, W.~B. 2007, \mnras, 380, 828

\bibitem[{{Jorstad} \& {Marscher}(2006)}]{jorstad2006}
{Jorstad}, S.~G., \& {Marscher}, A.~P. 2006, Astronomische Nachrichten, 327,
  227

\bibitem[{{Jorstad} {et~al.}(2005){Jorstad}, {Marscher}, {Lister}, {Stirling},
  {Cawthorne}, {Gear}, {G{\'o}mez}, {Stevens}, {Smith}, {Forster}, \&
  {Robson}}]{jorstad2005}
{Jorstad}, S.~G., {Marscher}, A.~P., {Lister}, M.~L., {et~al.} 2005, \aj, 130,
  1418

\bibitem[{{Kadler} {et~al.}(2016){Kadler}, {Krau{\ss}}, {Mannheim}, {Ojha},
  {M{\"u}ller}, {Schulz}, {Anton}, {Baumgartner}, {Beuchert}, {Buson},
  {Carpenter}, {Eberl}, {Edwards}, {Eisenacher Glawion}, {Els{\"a}sser},
  {Gehrels}, {Gr{\"a}fe}, {Gulyaev}, {Hase}, {Horiuchi}, {James}, {Kappes},
  {Kappes}, {Katz}, {Kreikenbohm}, {Kreter}, {Kreykenbohm}, {Langejahn},
  {Leiter}, {Litzinger}, {Longo}, {Lovell}, {McEnery}, {Natusch}, {Phillips},
  {Pl{\"o}tz}, {Quick}, {Ros}, {Stecker}, {Steinbring}, {Stevens}, {Thompson},
  {Tr{\"u}stedt}, {Tzioumis}, {Weston}, {Wilms}, \& {Zensus}}]{kadler}
{Kadler}, M., {Krau{\ss}}, F., {Mannheim}, K., {et~al.} 2016, Nature Physics,
  12, 807

\bibitem[{{Kataoka} \& {Stawarz}(2005{\natexlab{a}})}]{kataoka2005}
{Kataoka}, J., \& {Stawarz}, {\L}. 2005{\natexlab{a}}, \apj, 622, 797

\bibitem[{{Kataoka} \& {Stawarz}(2005{\natexlab{b}})}]{kataoka06}
---. 2005{\natexlab{b}}, \apj, 622, 797

\bibitem[{{Kharb} {et~al.}(2012){Kharb}, {Lister}, {Marshall}, \&
  {Hogan}}]{kharb2012}
{Kharb}, P., {Lister}, M.~L., {Marshall}, H.~L., \& {Hogan}, B.~S. 2012, \apj,
  748, 81

\bibitem[{{Kim} {et~al.}(2015){Kim}, {Im}, {Kim}, {Jun}, {Woo}, {Lee}, {Lee},
  {Nakagawa}, {Matsuhara}, {Wada}, {Oyabu}, {Takagi}, {Ohyama}, \&
  {Lee}}]{kim2015}
{Kim}, D., {Im}, M., {Kim}, J.~H., {et~al.} 2015, \apjs, 216, 17

\bibitem[{{Kusunose} \& {Takahara}(2017)}]{kusunose17}
{Kusunose}, M., \& {Takahara}, F. 2017, \apj, 835, 20

\bibitem[{{Lister} {et~al.}(2009){Lister}, {Cohen}, {Homan}, {Kadler},
  {Kellermann}, {Kovalev}, {Ros}, {Savolainen}, \& {Zensus}}]{lister2009}
{Lister}, M.~L., {Cohen}, M.~H., {Homan}, D.~C., {et~al.} 2009, \aj, 138, 1874

\bibitem[{{Lister} {et~al.}(2016){Lister}, {Aller}, {Aller}, {Homan},
  {Kellermann}, {Kovalev}, {Pushkarev}, {Richards}, {Ros}, \&
  {Savolainen}}]{lister2016}
{Lister}, M.~L., {Aller}, M.~F., {Aller}, H.~D., {et~al.} 2016, \aj, 152, 12

\bibitem[{{Liu} {et~al.}(2006){Liu}, {Jiang}, \& {Gu}}]{liu2006}
{Liu}, Y., {Jiang}, D.~R., \& {Gu}, M.~F. 2006, \apj, 637, 669

\bibitem[{{Lovell} {et~al.}(2000){Lovell}, {Tingay}, {Piner}, {Jauncey},
  {Preston}, {Murphy}, {McCulloch}, {Costa}, {Nicolson}, {Hirabayashi},
  {Reynolds}, {Tzioumis}, {Jones}, {Lister}, {Meier}, {Birkinshaw}, {Chartas},
  {Feigleson}, {Garmire}, {Ghosh}, {Marshall}, {Mathur}, {Sambruna},
  {Schwartz}, {Tucker}, {Wilkes}, \& {Worrall}}]{lovell2000}
{Lovell}, J.~E.~J., {Tingay}, S.~J., {Piner}, B.~G., {et~al.} 2000, in
  Astrophysical Phenomena Revealed by Space VLBI, ed. H.~{Hirabayashi}, P.~G.
  {Edwards}, \& D.~W. {Murphy}, 215--218

\bibitem[{{Mannheim} \& {Biermann}(1989)}]{mannheim89}
{Mannheim}, K., \& {Biermann}, P.~L. 1989, \aap, 221, 211

\bibitem[{{Marshall} {et~al.}(2011){Marshall}, {Gelbord}, {Schwartz}, {Murphy},
  {Lovell}, {Worrall}, {Birkinshaw}, {Perlman}, {Godfrey}, \&
  {Jauncey}}]{marshall2011}
{Marshall}, H.~L., {Gelbord}, J.~M., {Schwartz}, D.~A., {et~al.} 2011, \apjs,
  193, 15

\bibitem[{{Martel} {et~al.}(2003){Martel}, {Ford}, {Tran}, {Illingworth},
  {Krist}, {White}, {Sparks}, {Gronwall}, {Cross}, {Hartig}, {Clampin},
  {Ardila}, {Bartko}, {Ben{\'{\i}}tez}, {Blakeslee}, {Bouwens}, {Broadhurst},
  {Brown}, {Burrows}, {Cheng}, {Feldman}, {Franx}, {Golimowski}, {Infante},
  {Kimble}, {Lesser}, {McCann}, {Menanteau}, {Meurer}, {Miley}, {Postman},
  {Rosati}, {Sirianni}, {Tsvetanov}, \& {Zheng}}]{martel03}
{Martel}, A.~R., {Ford}, H.~C., {Tran}, H.~D., {et~al.} 2003, \aj, 125, 2964

\bibitem[{{McNamara} {et~al.}(2009){McNamara}, {Kuncic}, \&
  {Wu}}]{McNamara2009}
{McNamara}, A.~L., {Kuncic}, Z., \& {Wu}, K. 2009, \mnras, 395, 1507

\bibitem[{{Mertens} \& {Lobanov}(2015)}]{mertens2015_wise}
{Mertens}, F., \& {Lobanov}, A. 2015, \aap, 574, A67

\bibitem[{{Mertens} \& {Lobanov}(2016)}]{mertens2016}
{Mertens}, F., \& {Lobanov}, A.~P. 2016, \aap, 587, A52

\bibitem[{{Meyer} {et~al.}(2017){Meyer}, {Breiding}, {Georganopoulos}, {Oteo},
  {Zwaan}, {Laing}, {Godfrey}, \& {Ivison}}]{meyer17}
{Meyer}, E.~T., {Breiding}, P., {Georganopoulos}, M., {et~al.} 2017, \apjl,
  835, L35

\bibitem[{{Meyer} {et~al.}(2011){Meyer}, {Fossati}, {Georganopoulos}, \&
  {Lister}}]{meyer2011}
{Meyer}, E.~T., {Fossati}, G., {Georganopoulos}, M., \& {Lister}, M.~L. 2011,
  \apj, 740, 98

\bibitem[{{Meyer} {et~al.}(2012){Meyer}, {Fossati}, {Georganopoulos}, \&
  {Lister}}]{meyer12}
---. 2012, \apjl, 752, L4

\bibitem[{{Meyer} \& {Georganopoulos}(2014)}]{meyer2014}
{Meyer}, E.~T., \& {Georganopoulos}, M. 2014, \apjl, 780, L27

\bibitem[{{Meyer} {et~al.}(2015){Meyer}, {Georganopoulos}, {Sparks}, {Godfrey},
  {Lovell}, \& {Perlman}}]{meyer2015}
{Meyer}, E.~T., {Georganopoulos}, M., {Sparks}, W.~B., {et~al.} 2015, \apj,
  805, 154

\bibitem[{{Meyer} {et~al.}(2016){Meyer}, {Sparks}, {Georganopoulos},
  {Anderson}, {van der Marel}, {Biretta}, {Sohn}, {Chiaberge}, {Perlman}, \&
  {Norman}}]{meyer16_hst}
{Meyer}, E.~T., {Sparks}, W.~B., {Georganopoulos}, M., {et~al.} 2016, \apj,
  818, 195

\bibitem[{{Miller} {et~al.}(2006){Miller}, {Brandt}, {Gallagher}, {Laor},
  {Wills}, {Garmire}, \& {Schneider}}]{miller06}
{Miller}, B.~P., {Brandt}, W.~N., {Gallagher}, S.~C., {et~al.} 2006, \apj, 652,
  163

\bibitem[{{Mullin} \& {Hardcastle}(2009)}]{mullin2009}
{Mullin}, L.~M., \& {Hardcastle}, M.~J. 2009, \mnras, 398, 1989

\bibitem[{{Perlman} {et~al.}(2011){Perlman}, {Georganopoulos}, {Marshall},
  {Schwartz}, {Padgett}, {Gelbord}, {Lovell}, {Worrall}, {Birkinshaw},
  {Murphy}, \& {Jauncey}}]{perlman11}
{Perlman}, E.~S., {Georganopoulos}, M., {Marshall}, H.~L., {et~al.} 2011, \apj,
  739, 65

\bibitem[{{Petropoulou} {et~al.}(2017){Petropoulou}, {Vasilopoulos}, \&
  {Giannios}}]{petropoulou2017}
{Petropoulou}, M., {Vasilopoulos}, G., \& {Giannios}, D. 2017, \mnras, 464,
  2213

\bibitem[{{Sambruna} {et~al.}(2004){Sambruna}, {Gambill}, {Maraschi},
  {Tavecchio}, {Cerutti}, {Cheung}, {Urry}, \& {Chartas}}]{sambruna2004}
{Sambruna}, R.~M., {Gambill}, J.~K., {Maraschi}, L., {et~al.} 2004, \apj, 608,
  698

\bibitem[{{Sambruna} {et~al.}(2006){Sambruna}, {Gliozzi}, {Donato}, {Maraschi},
  {Tavecchio}, {Cheung}, {Urry}, \& {Wardle}}]{sambruna06}
{Sambruna}, R.~M., {Gliozzi}, M., {Donato}, D., {et~al.} 2006, \apj, 641, 717

\bibitem[{{Sambruna} {et~al.}(2002){Sambruna}, {Maraschi}, {Tavecchio}, {Urry},
  {Cheung}, {Chartas}, {Scarpa}, \& {Gambill}}]{sambruna02}
{Sambruna}, R.~M., {Maraschi}, L., {Tavecchio}, F., {et~al.} 2002, \apj, 571,
  206

\bibitem[{{Schwartz} {et~al.}(2000){Schwartz}, {Marshall}, {Lovell}, {Piner},
  {Tingay}, {Birkinshaw}, {Chartas}, {Elvis}, {Feigelson}, {Ghosh}, {Harris},
  {Hirabayashi}, {Hooper}, {Jauncey}, {Lanzetta}, {Mathur}, {Preston},
  {Tucker}, {Virani}, {Wilkes}, \& {Worrall}}]{schwartz2000}
{Schwartz}, D.~A., {Marshall}, H.~L., {Lovell}, J.~E.~J., {et~al.} 2000, \apjl,
  540, 69

\bibitem[{{Stanley} {et~al.}(2015){Stanley}, {Kharb}, {Lister}, {Marshall},
  {O'Dea}, \& {Baum}}]{stanley15}
{Stanley}, E.~C., {Kharb}, P., {Lister}, M.~L., {et~al.} 2015, \apj, 807, 48

\bibitem[{{Stawarz} {et~al.}(2004){Stawarz}, {Sikora}, {Ostrowski}, \&
  {Begelman}}]{stawarz2004}
{Stawarz}, {\L}., {Sikora}, M., {Ostrowski}, M., \& {Begelman}, M.~C. 2004,
  \apj, 608, 95

\bibitem[{{Tavecchio} {et~al.}(2003){Tavecchio}, {Ghisellini}, \&
  {Celotti}}]{tavecchio03}
{Tavecchio}, F., {Ghisellini}, G., \& {Celotti}, A. 2003, \aap, 403, 83

\bibitem[{{Tavecchio} {et~al.}(2000){Tavecchio}, {Maraschi}, {Sambruna}, \&
  {Urry}}]{tavecchio2000}
{Tavecchio}, F., {Maraschi}, L., {Sambruna}, R.~M., \& {Urry}, C.~M. 2000,
  \apjl, 544, L23

\bibitem[{{Tavecchio} {et~al.}(2007){Tavecchio}, {Maraschi}, {Wolter},
  {Cheung}, {Sambruna}, \& {Urry}}]{tavecchio2007}
{Tavecchio}, F., {Maraschi}, L., {Wolter}, A., {et~al.} 2007, \apj, 662, 900

\bibitem[{{Uchiyama}(2008)}]{uchiyama2008}
{Uchiyama}, Y. 2008, International Journal of Modern Physics D, 17, 1475

\bibitem[{{Uchiyama} {et~al.}(2006){Uchiyama}, {Urry}, {Cheung}, {Jester}, {Van
  Duyne}, {Coppi}, {Sambruna}, {Takahashi}, {Tavecchio}, \&
  {Maraschi}}]{uchiyama2006}
{Uchiyama}, Y., {Urry}, C.~M., {Cheung}, C.~C., {et~al.} 2006, \apj, 648, 910

\bibitem[{{Wilson} \& {Yang}(2002)}]{wilson2002}
{Wilson}, A.~S., \& {Yang}, Y. 2002, \apj, 568, 133

\end{thebibliography}

\acknowledgments 
E.T.M. acknowledges HST Grant GO-13676. 

E.T.M. and M.G. acknowledge NASA Grant NNX15AE55G.

This research has made use of data from the MOJAVE database that is maintained by the MOJAVE team \citep{lister2009}.

This paper makes use of the following ALMA data: ADS/JAO.ALMA\#2015.1.00932.S, ADS/JAO.ALMA\#2016.1.01481.S, and ADS/JAO.ALMA\#2016.1.01250.S. ALMA is a partnership of ESO (representing its member states), NSF (USA) and NINS (Japan), together with NRC (Canada), NSC and ASIAA (Taiwan), and KASI (Republic of Korea), in cooperation with the Republic of Chile. The Joint ALMA Observatory is operated by ESO, AUI/NRAO and NAOJ.

The National Radio Astronomy Observatory is a facility of the National Science Foundation operated under cooperative agreement by Associated Universities, Inc.

\end{document}